\documentstyle[amssymb,preprint,prb,aps]{revtex}

\begin{document}

\begin{titlepage}

\title
{Spin tunneling of trigonal and hexagonal ferromagnets in an arbitrarily
directed magnetic field}

\author{Rong L\"{u}\footnote {Author to whom 
the correspondence should be addressed.\\
Electronic address: rlu@castu.phys.tsinghua.edu.cn}, Jia-Lin Zhu, Zhi-Rong Liu,
Lee Chang, and Bing-Lin Gu} 
\address{Center for Advanced Study, Tsinghua University,
Beijing 100084, P. R. China}

\maketitle
\begin{abstract}
The quantum
tunneling of the magnetization vector are studied theoretically in
single-domain ferromagnetic nanoparticles
placed in an external magnetic field at an arbitrarily directed angle in
the  $ZX$ plane. We consider the magnetocrystalline anisotropy
with trigonal and hexagonal crystal symmetry, respectively.
By applying the instanton technique in the
spin-coherent-state path-integral representation, we
calculate the tunnel splittings, the
tunneling rates and the crossover temperatures in the low barrier
limit for different angle ranges of the external magnetic field
($\theta_{H}=\pi/2$, $\pi/2\ll\theta_{H}\ll\pi$, and $\theta_{H}=\pi$).
Our results show that the tunnel splittings, the
tunneling rates and the crossover temperatures depend on the 
orientation of the external magnetic
field distinctly, which provides a possible experimental
test for magnetic quantum tunneling in
nanometer-scale single-domain ferromagnets.

\noindent
{\bf PACS number(s)}:  75.45.+j, 75.50.Ee
\end{abstract}

\end{titlepage}

\section*{I. Introduction}

Recently, there has been great experimental and theoretical effort to
observe and interpret macroscopic quantum tunneling (MQT) and coherence
(MQC) in nanometer-scale magnets at sufficiently low temperature.\cite{1}
Theoretical investigations based on the spin-coherent-state path integral
were performed for the single-domain ferromagnetic (FM) nanoparticles, which
showed that MQT and MQC were possible in magnets containing as much as $%
10^5-10^6$ spins. Several experiments involving resonance measurements,
magnetic relaxation, and hysteresis loop study for various systems showed
either temperature-independent relaxation phenomena or a well-defined
resonance depending exponentially on the number of total spins, which
supported the idea of magnetic quantum tunneling.\cite{1}

More recently, the tunneling behaviors of the magnetization vector were
studied extensively for the single-domain FM nanoparticles in the presence
of an external magnetic field applied at an arbitrary angle. The MQT problem
for FM particles with uniaxial crystal symmetry was first studied by
Zaslavskii who calculated the tunneling exponent, the preexponential factors
and their temperature dependences in the low barrier limit with the help of
mapping the spin system onto a one-dimensional particle system.\cite{2} For
the same crystal symmetry, Miguel and Chudnovsky\cite{3} calculated the
tunneling rate by applying the imaginary-time path integral, and
demonstrated that the angular and field dependences of the tunneling
exponent obtained by Zaslavskii's method and by the path-integral method
coincide precisely. They also discussed the tunneling rate at finite
temperature and suggested experimental procedures.\cite{3} Kim and Hwang
performed a calculation based on the instanton technique for FM particles
with biaxial and tetragonal crystal symmetry,\cite{4} and Kim extended the
tunneling rate for biaxial crystal symmetry to a finite temperature.\cite{5}
The quantum-classical transition of the escape rate for FM particles with
uniaxial crystal symmetry in an arbitrarily directed field was investigated
by Garanin, Hidalgo and Chudnovsky with the help of mapping onto a particle
moving in a double-well potential.\cite{6} The switching field measurement
was carried out on single-domain FM nanoparticles of Barium ferrite
(BaFeCoTiO) containing about $10^5-10^6$ spins.\cite{7} The measured angular
dependance of the crossover temperature was found to be in excellent
agreement with the theoretical prediction,\cite{3} which strongly suggests
the MQT of magnetization in the BaFeCoTiO nanoparticles. L\"{u} {\it et al}.
studied the MQT and MQC of the N\'{e}el vector in single-domain
antiferromagnetic (AFM) nanoparticles with biaxial, tetragonal, and
hexagonal crystal symmetry in an arbitrarily directed field.\cite{8}

In this paper, we extend the previous theoretical results obtained for the
single-domain FM particles with biaxial and tetragonal symmetry to those for
FM particles with a much more complex structure placed in an external
magnetic field at an arbitrarily directed angle in the $ZX$ plane, based on
the instanton technique in the spin-coherent-state path-integral
representation. We consider the magnetocrystalline anisotropies with
trigonal and hexagonal crystal symmetry, respectively. Both the
Wentzel-Kramers-Brillouin (WKB) exponents and the preexponential factors are
evaluated analytically in the tunneling rates for MQT and the tunnel
splittings for MQC in FM particles for different angle ranges of the
external magnetic field $(\theta _H=\pi /2$, $\pi /2+O\left( \epsilon
^{3/2}\right) \ <\theta _H<\pi -O(\epsilon ^{3/2})$, and $\theta _H=\pi ),$
and the temperature which corresponds to the crossover from the thermal to
the quantum regime is clearly shown for each case. Our results show that the
distinct angular dependence, together with the dependence of the WKB
tunneling rate and the crossover temperature on the strength of the external
magnetic field, may provide an independent experimental test for the
magnetic tunneling in single-domain FM nanoparticles. The calculations
performed in this paper are semiclassical in nature, i.e., valid for large
spins and in the continuum limit. We analyze the validity of the
semiclassical approximation, and find that the semiclassical approximation
is rather good for the typical values of parameters for single-domain FM
nanoparticles.

This paper is structured in the following way. In Sec. II, we briefly review
the basic ideas of the MQT and MQC in single-domain FM particles. In Secs.
III and IV, we study the quantum tunneling of the magnetization vector for
FM particles with trigonal and hexagonal crystal symmetry in the presence of
an external magnetic field applied in the $ZX$ plane with a range of angles $%
\pi /2\leq \theta _H\leq \pi $. The conclusions are presented in Sec. V. In
Appendix A, we explain briefly the computation of the preexponential factors
\/in the WKB tunneling rate, and then apply this approach to obtain the
tunnel splittings for FM particles with trigonal crystal symmetry in a
magnetic field applied perpendicular to the anisotropy axis $(\theta _H=\pi
/2)$ in detail.

\section*{II. MQT and MQC of the magnetization vector in FM particles}

In this section we briefly review some basic ideas of MQT and MQC of the
magnetization vector in single-domain FM nanoparticles, based on the
instanton technique in the spin-coherent-state path integral.\cite{1,11,12}

The system of interest is a nanometer-scale single-domain ferromagnet at a
temperature well below its anisotropy gap. For such a FM particle, the
tunnel splitting for MQC or the tunneling rate for MQT is determined by the
imaginary-time transition amplitude from an initial state $\left|
i\right\rangle $ to a final state $\left| f\right\rangle $ as 
\begin{equation}
U_{fi}=\left\langle f\right| e^{-HT}\left| i\right\rangle =\int D\Omega \exp
\left( -S_E\right) ,  \eqnum{1}
\end{equation}
where $S_E$ is the Euclidean action and $D\Omega $ is the measurement of the
path integral. In the spin-coherent-state path-integral representation, the
Euclidean action can be expressed as 
\begin{equation}
S_E\left( \theta ,\phi \right) =\frac V\hbar \int d\tau \left[ i\frac{M_0}%
\gamma \left( \frac{d\phi }{d\tau }\right) -i\frac{M_0}\gamma \left( \frac{%
d\phi }{d\tau }\right) \cos \theta +E\left( \theta ,\phi \right) \right] , 
\eqnum{2}
\end{equation}
where $V$ is the volume of the FM particle and $\gamma $ is the gyromagnetic
ratio. $M_0=\left| \overrightarrow{M}\right| =\hbar \gamma S/V$, where $S$
is the total spin of FM particles. It is noted that the first two terms in
Eq. (2) define the topological Berry or Wess-Zumino, Chern-Simons term which
arises from the nonorthogonality of spin coherent states. The Wess-Zumino
term has a simple topological interpretation. For a closed path, this term
equals $-iS$ times the area swept out on the unit sphere between the path
and the north pole. The first term in Eq. (2) is a total imaginary-time
derivative, which has no effect on the classical equations of motion, but it
is crucial for the spin-parity effects.\cite{9,10} However, for the closed
instanton or bounce trajectory described in this paper (as shown in the
following), this time derivative gives a zero contribution to the path
integral, and therefore can be omitted.

In discussing macroscopic quantum phenomena, it is essential to distinguish
between two types of processes: MQC (i.e., coherent tunneling) and MQT
(i.e., incoherent tunneling). In the case of MQC, the system in question
performs coherent NH$_3$-type oscillations between two degenerate wells
separated by a classically impenetrable barrier. Tunneling between
neighboring degenerate vacua can be described by the instanton configuration
with nonzero topological charge and leads to a level splitting of the ground
states.\cite{11} The tunneling removes the degeneracy of the original ground
states, and the true ground state is a superposition of the previous
degenerate ground states. For the case of MQT, the system escapes from a
metastable potential well into a continuum by quantum tunneling at
sufficiently low temperatures, and the tunneling results in an imaginary
part of the energy which is dominated by the so-called bounce configuration
with zero topological charge.\cite{11} As emphasized by Leggett, the two
phenomena of MQC and MQT are physically very different, particularly from
the viewpoint of experimental feasibility.\cite{20} MQC is a far more
delicate phenomenon than MQT, as it is much more easily destroyed by an
environment,\cite{21} and by very small $c$-number symmetry breaking fields
that spoil the degeneracy.

In the semiclassical limit, the dominant contribution to the transition
amplitude comes from the finite action solution (instanton) of the classical
equation of motion. The motion of the magnetization vector $\overrightarrow{M%
}$ is determined by the Landau-Lifshitz equation, 
\begin{equation}
i\frac{d\overrightarrow{M}}{d\tau }=-\gamma \overrightarrow{M}\times \frac{%
dE\left( \overrightarrow{M}\right) }{d\overrightarrow{M}},  \eqnum{3}
\end{equation}
which can also be expressed as the following equations in the spherical
coordinate system, 
\begin{eqnarray}
i\left( \frac{d\overline{\theta }}{d\tau }\right) \sin \overline{\theta } &=&%
\frac \gamma {M_0}\frac{\partial E}{\partial \overline{\phi }},  \eqnum{4a}
\\
i\left( \frac{d\overline{\phi }}{d\tau }\right) \sin \overline{\theta } &=&-%
\frac \gamma {M_0}\frac{\partial E}{\partial \overline{\theta }},  \eqnum{4b}
\end{eqnarray}
where $\overline{\theta }$ and $\overline{\phi }$ denote the classical path.
Note that the Euclidean action Eq. (2) describes the $\left( 1\oplus
1\right) $-dimensional dynamics in the Hamiltonian formulation with
canonical variables $\phi $ and $P_\phi =S(1-$cos$\theta )$. The instanton's
contribution to the tunneling rate $\Gamma $ for MQT or the tunnel splitting 
$\Delta $ for MQC (not including the topological Wess-Zumino or Berry phase)
is given by\cite{11,12} 
\begin{equation}
\Gamma \ (\text{or }\ \Delta )=A\omega _p\left( \frac{S_{cl}}{2\pi }\right)
^{1/2}e^{-S_{cl}},  \eqnum{5}
\end{equation}
where $\omega _p$ is the frequency of small oscillations near the bottom of
the inverted potential, and $S_{cl}$ is the classical action. The
preexponential factor $A$ originates from the quantum fluctuations about the
classical path, which can be evaluated by expanding the Euclidean action to
second order in the small fluctuations.\cite{11,12} In Ref. 12, Garg and Kim
studied the general formalism for calculating both the exponent and the
preexponential factors in the WKB tunneling rates for MQT and MQC in
single-domain FM nanoparticles. In Appendix A, we explain briefly the basic
idea of this calculation, and then apply this approach to calculate the
instanton's contribution to the tunnel splittings for MQC of the
magnetization vector in FM particles with trigonal crystal symmetry in an
external magnetic field perpendicular to the anisotropy axis (considered in
Sec. III) in detail.

\section*{III. MQC and MQT for trigonal crystal symmetry}

In this section, we study the tunneling behaviors of the magnetization
vector in single-domain FM nanoparticle with trigonal crystal symmetry. The
external magnetic field is applied in the $ZX$ plane, at an angle in the
range of $\pi /2\leq \theta _H<\pi $. Now the total energy $E\left( \theta
,\phi \right) $ can be written as 
\begin{equation}
E\left( \theta ,\phi \right) =K_1\sin ^2\theta -K_2\sin ^3\theta \cos \left(
3\phi \right) -M_0H_x\sin \theta \cos \phi -M_0H_z\cos \theta +E_0, 
\eqnum{6}
\end{equation}
where $K_1$ and $K_2$ are the magnetic anisotropy constants satisfying $%
K_1\gg K_2>0$, and $E_0$ is a constant which makes $E\left( \theta ,\phi
\right) $ zero at the initial orientation. As the magnetic field is applied
in the $ZX$ plane, $H_x=H\sin \theta _H$ and $H_z=H\cos \theta _H$, where $H$
is the magnitude of the field and $\theta _H$ is the angle between the
magnetic field and the $\widehat{z}$ axis.

In the absence of the external magnetic field, the system reduces to one
with threefold rotational symmetry around $\widehat{z}$ axis and reflection
symmetry in the $XY$ plane. The unit vectors $\widehat{z}$ and $-\widehat{z}$
define the two classical ground state configurations. The transition
amplitude between degenerate ground states can be suppressed to zero
resulting from the destructive Wess-Zumino phase if the system has
time-reversal invariance at zero magnetic field.\cite{10} However, for the
closed instanton or bounce trajectory described in this paper (as shown in
the following) the phase term in Eq. (2), proportional to $d\phi /d\tau $
(not $\left( d\phi /d\tau \right) \cos \theta $ term) gives a zero
contribution to the integral Eq. (2) and, therefore, can be omitted.

By introducing the dimensionless parameters as 
\begin{equation}
\overline{K}_2=K_2/2K_1,\overline{H}_x=H_x/H_0,\overline{H}_z=H_z/H_0, 
\eqnum{7}
\end{equation}
Eq. (6) can be rewritten as 
\begin{equation}
\overline{E}\left( \theta ,\phi \right) =\frac 12\sin ^2\theta -\overline{K}%
_2\sin ^3\theta \cos \left( 3\phi \right) -\overline{H}_x\sin \theta \cos
\phi -\overline{H}_z\cos \theta +\overline{E}_0,  \eqnum{8}
\end{equation}
where $E\left( \theta ,\phi \right) =2K_1\overline{E}\left( \theta ,\phi
\right) $, and $H_0=2K_1/M_0$. At finite magnetic field, the plane given by $%
\phi =0$ is the easy plane, on which $\overline{E}\left( \theta ,\phi
\right) $ reduces to 
\begin{equation}
\overline{E}\left( \theta ,\phi =0\right) =\frac 12\sin ^2\theta -\overline{K%
}_2\sin ^3\theta -\overline{H}\cos \left( \theta -\theta _H\right) +%
\overline{E}_0.  \eqnum{9}
\end{equation}
We denote $\theta _0$ to be the initial angle and $\theta _c$ the critical
angle at which the energy barrier vanishes when the external magnetic field
is close to the critical value $\overline{H}_c\left( \theta _H\right) $ (to
be calculated in the following). Then, the initial angle $\theta _0$
satisfies $\left[ d\overline{E}\left( \theta ,\phi =0\right) /d\theta
\right] _{\theta =\theta _0}=0$, the critical angle $\theta _c$ and the
dimensionless critical field $\overline{H}_c$ satisfy both $\left[ d%
\overline{E}\left( \theta ,\phi =0\right) /d\theta \right] _{\theta =\theta
_c,\overline{H}=\overline{H}_c}=0$ and $\left[ d^2\overline{E}\left( \theta
,\phi =0\right) /d\theta ^2\right] _{\theta =\theta _c,\overline{H}=%
\overline{H}_c}=0$, which leads to 
\begin{eqnarray}
\frac 12\sin \left( 2\theta _0\right) -3\overline{K}_2\sin ^2\theta _0\cos
\theta _0+\overline{H}\sin \left( \theta _0-\theta _H\right) &=&0, 
\eqnum{10a} \\
\frac 12\sin \left( 2\theta _c\right) -3\overline{K}_2\sin ^2\theta _c\cos
\theta _c+\overline{H}_c\sin \left( \theta _c-\theta _H\right) &=&0, 
\eqnum{10b} \\
\cos \left( 2\theta _c\right) -3\overline{K}_2\left( 2\sin \theta _c\cos
^2\theta _c-\sin ^3\theta _c\right) +\overline{H}_c\cos \left( \theta
_c-\theta _H\right) &=&0.  \eqnum{10c}
\end{eqnarray}
After some algebra, $\overline{H}_c\left( \theta _H\right) $ and $\theta _c$
are found to be 
\begin{eqnarray}
\overline{H}_c &=&\frac 1{\left[ \left( \sin \theta _H\right) ^{2/3}+\left|
\cos \theta _H\right| ^{2/3}\right] ^{3/2}}\left[ 1+3\overline{K}_2\frac 1{%
\left( 1+\left| \cot \theta _H\right| ^{2/3}\right) ^{1/2}}\right.  \nonumber
\\
&&\left. +6\overline{K}_2\frac 1{\left( 1+\left| \cot \theta _H\right|
^{2/3}\right) ^{3/2}}\right] ,  \eqnum{11a} \\
\sin ^2\theta _c &=&\frac 1{1+\left| \cot \theta _H\right| ^{2/3}}\left[ 1-2%
\overline{K}_2\frac{\left| \cot \theta _H\right| ^{2/3}}{\left( 1+\left|
\cot \theta _H\right| ^{2/3}\right) ^{1/2}}-4\overline{K}_2\frac{\left| \cot
\theta _H\right| ^{2/3}}{\left( 1+\left| \cot \theta _H\right| ^{2/3}\right)
^{3/2}}\right] .  \eqnum{11b}
\end{eqnarray}

Now we consider the limiting case that the external magnetic field is
slightly lower than the critical field, i.e., $\epsilon =1-\overline{H}/%
\overline{H}_c\ll 1$. At this practically interesting situation, the barrier
height is low and the width is narrow, and therefore the tunneling rate in
MQT or the tunnel splitting in MQC is large. Introducing $\eta \equiv \theta
_c-\theta _0$ $\left( \left| \eta \right| \ll 1\text{ in the limit of }%
\epsilon \ll 1\right) $, expanding $\left[ d\overline{E}\left( \theta ,\phi
=0\right) /d\theta \right] _{\theta =\theta _0}=0$ about $\theta _c$, and
using the relations $\left[ d\overline{E}\left( \theta ,\phi =0\right)
/d\theta \right] _{\theta =\theta _c,\overline{H}=\overline{H}_c}=0$ and $%
\left[ d^2\overline{E}\left( \theta ,\phi =0\right) /d\theta ^2\right]
_{\theta =\theta _c,\overline{H}=\overline{H}_c}=0$, we obtain the
approximation equation for $\eta $ in the order of $\epsilon ^{3/2}$, 
\begin{eqnarray}
&&\left. -\epsilon \overline{H}_c\sin \left( \theta _c-\theta _H\right)
-\eta ^2\left( \frac 34\sin 2\theta _c+3\overline{K}_2\cos 3\theta _c\right)
\right.  \nonumber \\
&&\left. +\eta \left[ \epsilon \overline{H}_c\cos \left( \theta _c-\theta
_H\right) +\eta ^2\left( \frac 12\cos 2\theta _c-3\overline{K}_2\sin 3\theta
_c\right) \right] =0.\right.  \eqnum{12}
\end{eqnarray}
Then $\overline{E}\left( \theta ,\phi \right) $ reduces to the following
equation in the limit of small $\epsilon $, 
\begin{equation}
\overline{E}\left( \delta ,\phi \right) =2\overline{K}_2\sin ^2\left( 3\phi
/2\right) \sin ^3\left( \theta _0+\delta \right) +\overline{H}_x\sin \left(
\theta _0+\delta \right) \left( 1-\cos \phi \right) +\overline{E}_1\left(
\delta \right) ,  \eqnum{13}
\end{equation}
where $\delta \equiv \theta -\theta _0$ $\left( \left| \delta \right| \ll 1%
\text{ in the limit of }\epsilon \ll 1\right) $, and $\overline{E}_1\left(
\delta \right) $ is a function of only $\delta $ given by 
\begin{eqnarray}
\overline{E}_1\left( \delta \right) &=&-\frac 12\left[ \overline{H}_c\sin
\left( \theta _c-\theta _H\right) -\overline{K}_2\left( \cos ^3\theta _c-%
\frac 32\sin ^2\theta _c\cos \theta _c\right) \right] \left( 3\delta ^2\eta
-\delta ^3\right)  \nonumber \\
&&-\frac 12\left[ \overline{H}_c\cos \left( \theta _c-\theta _H\right) -3%
\overline{K}_2\left( \sin ^3\theta _c-4\sin \theta _c\cos ^2\theta _c\right)
\right] \left[ \delta ^2\left( \epsilon -\frac 32\eta ^2\right) +\delta
^3\eta -\frac 14\delta ^4\right]  \nonumber \\
&&-\frac 32\overline{K}_2\left( \sin ^3\theta _c-4\sin \theta _c\cos
^2\theta _c\right) \delta ^2\epsilon .  \eqnum{14}
\end{eqnarray}

In the following, we will investigate the tunneling behaviors of the
magnetization vector in FM particles with trigonal crystal symmetry at
different angle ranges of the external magnetic field as $\theta _H=\pi /2$
and $\pi /2<\theta _H<\pi $, respectively.

\subsection*{A. $\theta _H=\pi /2$}

For $\theta _H=\pi /2$, we have $\theta _c=\pi /2$ from Eq. (11b) and $\eta =%
\sqrt{2\epsilon }\left( 1+9\overline{K}_2/2\right) $ from Eq. (12). Then $%
\overline{E}_1\left( \delta \right) $ of Eq. (14) reduces to 
\begin{equation}
\overline{E}_1\left( \delta \right) =\frac 18\delta ^2\left( \delta -2\eta
\right) ^2.  \eqnum{15}
\end{equation}
The plot of the effective potential $\overline{E}_1\left( \delta \right) $
as a function of $\delta \left( =\theta -\theta _0\right) $ for $\theta
_H=\pi /2$ is shown in Fig. 1. Now the problem is one of MQC, where the
magnetization vector resonates coherently between the energetically
degenerate easy directions at $\delta =0$ and $\delta =2\sqrt{2\epsilon }%
\left( 1+9\overline{K}_2/2\right) $ separated by a classically impenetrable
barrier at $\delta =\sqrt{2\epsilon }\left( 1+9\overline{K}_2/2\right) $.
Substituting Eq. (15) into the classical equations of motion, we obtain the
classical solution called instanton as 
\begin{eqnarray}
\overline{\phi } &=&i\epsilon \left( 1+3\overline{K}_2+\frac 12\epsilon
\right) \frac 1{\cosh ^2\left( \overline{\omega }_c\overline{\tau }\right) },
\nonumber \\
\overline{\delta } &=&\sqrt{2\epsilon }\left( 1+\frac 92\overline{K}%
_2\right) \left[ 1+\tanh \left( \overline{\omega }_c\overline{\tau }\right)
\right] ,  \eqnum{16}
\end{eqnarray}
where $\overline{\omega }_c=\sqrt{\epsilon /2}\left( 1+21\overline{K}%
_2/2-\epsilon /2\right) $, $\overline{\tau }=\omega _0\tau $, and $\omega
_0=2K_1V/\hbar S$. We can calculate the classical action by integrating the
Euclidean action Eq. (2) with the above classical trajectory, and the result
is found to be 
\begin{equation}
S_{cl}=\frac{2^{5/2}}3\epsilon ^{3/2}S\left( 1+\frac{15}2\overline{K}_2+%
\frac 12\epsilon \right) .  \eqnum{17}
\end{equation}

Now we consider the transition exponent which is usually addressed by
experiments. Transitions between two states in a bistable system or escaping
from a metastable state can occur either due to the quantum tunneling or via
the classical thermal activation. In the limit of temperature $T\rightarrow
0 $, the transitions are purely quantum-mechanical and the rate goes as $%
\Gamma \sim \exp \left( -S_{cl}\right) $, with $S_{cl}$ being the classical
action or the WKB\ exponent which is independent of temperature. As the
temperature increases from zero, thermal effects enter in the quantum
tunneling process. If the temperature is sufficiently high, the decay from a
metastable state is determined by processes of thermal activation, and the
transition rate follows the Arrhenius law, $\Gamma \sim \exp \left(
-U/k_BT\right) $, with $k_B$ being the Boltzmann constant and $U$ being the
height of energy barrier between the two states. Because of the exponential
dependence of the thermal rate on $T$, the temperature $T_c$ characterizing
the crossover from quantum to thermal regime can be estimated as $%
k_BT_c=U/S_{cl}$. For a quasiparticle with the effective mass $M$ moving in
one-dimensional potential $U\left( x\right) $, a more accurate definition of
the crossover temperature in the absence of any dissipation was presented by
Goldanskii,\cite{13,14} $k_BT_c^{\prime }=\hbar \omega _b/2\pi $, where $%
\omega _b=\sqrt{-U^{\prime \prime }\left( x_b\right) /M}$ is the frequency
of small oscillations near the bottom of the inverted potential, $-U\left(
x\right) $, and $x_b$ corresponds to the bottom of inverted potential. Below 
$T_c^{\prime }$, thermally assisted quantum tunneling occurs from the
excited levels, that further reduces to the quantum tunneling from the
ground-state level as the temperature decreases to zero. Above $T_c^{\prime
} $, quantum tunneling effects are small and the transitions occur due to
the thermal activation to the top of the barrier. For the MQT problem, i.e.,
the problem of decay from the metastable state, both $T_c$ and $T_c^{\prime
} $ can be used as the definition of the crossover temperature corresponding
to the crossover from classical to quantum behavior since the quantum
escaping from a metastable state is one process of incoherent tunneling.
However, for the MQC problem, i.e., the problem of resonance between
degenerate states, the situation is different. As the temperature growing
from zero, three kinds of transitions should be taken into account: quantum
coherence between the degenerate ground-state levels (coherent tunneling),
quantum tunneling from the excited levels (thermally assisted tunneling or
incoherent tunneling), and classical over-barrier transition (incoherent
transition). Two kinds of crossover temperatures can be defined to
distinguish the three regimes. The Goldanskii definition $T_c^{\prime }$ for
MQC problem corresponds to the crossover from quantum coherence between the
degenerate ground-state levels (coherent tunneling) to quantum tunneling
from the excited levels (thermally assisted tunneling or incoherent
tunneling), while $T_c$ corresponds to the crossover from quantum coherence
between the degenerate ground-state levels (coherent tunneling) to classical
over-barrier transition (incoherent transition). Experiments involving
magnetic relaxation and resonance measurements for various systems have
shown either temperature-independent relaxation phenomena (in MQT) or a
well-defined resonance (in MQC) below some crossover temperature, which
strongly support the existence of quantum tunneling processes.\cite{1} And
more recently, the crossover from quantum to classical behavior and
associated phase transition have been investigated extensively in MQT and
MQC in single-domain FM particles.\cite{14,15,16,17,18} It is noted that the
sharpness of the crossover between thermal and quantum regimes also depends
on the strength of the dissipation with environment. In the case of the low
dissipation which is common for the magnetic systems, its effect on the
crossover is small.\cite{13,14}

For the single-domain FM nanoparticle in a magnetic field applied at $\theta
_H=\pi /2$, the magnetization vector resonates coherently between the
energetically degenerate easy directions at $\delta =0$ and $\delta =2\sqrt{%
2\epsilon }\left( 1+9\overline{K}_2/2\right) $ separated by a classically
impenetrable barrier at $\delta =\sqrt{2\epsilon }\left( 1+9\overline{K}%
_2/2\right) $, and the height of energy barrier is found to be: $%
U=K_1V\epsilon ^2\left( 1+18\overline{K}_2\right) $. Then, equating $S_{cl}$
to $U/k_BT$, we obtain that the crossover from quantum coherence between the
degenerate ground-state levels (coherent tunneling) to classical
over-barrier transition (incoherent transition) occurs at 
\begin{equation}
k_BT_c=\frac 3{2^{5/2}}\epsilon ^{1/2}\frac{K_1V}S\left( 1+\frac{21}2%
\overline{K}_2-\frac 12\epsilon \right) .  \eqnum{18}
\end{equation}
For this MQC problem, the Goldanskii definition $T_c^{\prime }$
corresponding the crossover from quantum coherence between the degenerate
ground-state levels (coherent tunneling) to quantum tunneling from excited
levels (thermally assisted tunneling or incoherent tunneling) becomes $%
k_BT_c^{\prime }=\hbar \omega _b/2\pi $, where $\omega _b=\overline{\omega }%
_b\omega _0$, with $\overline{\omega }_b\equiv \sqrt{-\overline{E}^{\prime
\prime }\left( \delta _m\right) /M}$ is the frequency of small oscillations
of the magnetization vector near the bottom of the inverted potential, $%
M^{-1}=\left( 1+12\overline{K}_2-\epsilon \right) $, and $\delta _m$ is the
position of the energy barrier. For the present case, $\delta _m=\sqrt{%
2\epsilon }\left( 1+9\overline{K}_2/2\right) $ and $\overline{\omega }_b=%
\sqrt{\epsilon }\left( 1+21\overline{K}_2/2-\epsilon /2\right) =\sqrt{2}%
\overline{\omega }_c$. Then it is easy to obtain that 
\begin{equation}
k_BT_c^{\prime }=\frac 1\pi \epsilon ^{1/2}\frac{K_1V}S\left( 1+\frac{21}2%
\overline{K}_2-\frac 12\epsilon \right) .  \eqnum{19}
\end{equation}
The comparison of Eqs. (18) and (19) shows that $T_c\approx 1.67T_c^{\prime
} $, which is consistent with the physical interpretation for
quantum-classical transition in the MQC problem.

It is noted that the quantum tunneling of the magnetization vector in
single-domain FM nanoparticles are studied with the help of the instanton
technique in the spin-coherent-state path-integral representation, which is
semiclassical in nature, i.e., valid for large spins and in the continuum
limit. Therefore, one should analyze the validity of the semiclassical
approximation. It is well known that for this approach to be valid, the
tunneling rate must be small, which indicates that the WKB exponent or the
classical action $S_{cl}\gg 1$. Moreover, the energy $\hbar \omega _b$ of
zero-point oscillations around the minimum of the inverted potential $-%
\overline{E}_1\left( \delta \right) $ should be sufficiently small compared
to the height of the barrier, $U=2K_1V\overline{E}_1\left( \delta _m\right) $
. For the single-domain FM nanoparticle with trigonal crystal symmetry in a
magnetic field applied at $\theta _H=\pi /2$, it is easy to show that the
WKB\ exponent is approximately given by 
\begin{equation}
B\backsim \frac U{\hbar \omega _b}=\frac 12\epsilon ^{3/2}S\left( 1+\frac{15}%
2\overline{K}_2+\frac 12\epsilon \right) ,  \eqnum{20}
\end{equation}
which agrees up to the numerical factor with the result of the classical
action in Eq. (17) obtained by applying the explicit instanton solution. For
the typical values of parameters for single-domain FM nanoparticles, $%
K_1\backsim 10^8$ erg/cm$^3$, $K_2\backsim 10^5$ erg/cm$^3$, and the total
spin $S=10^6$, we obtain that $B\backsim U/\hbar \omega _b\approx 15.8$ from
Eq. (20) and $S_{cl}\approx 59.6$ for $\epsilon =0.001$ from Eq. (17). In
this case the semiclassical approximation should be already rather good.

By applying the instanton technique for FM particles in the
spin-coherent-state path-integral representation,\cite{11,12} we obtain the
instanton's contribution to the tunnel splitting as (for detailed
calculation see Appendix A), 
\begin{equation}
\hbar \Delta _0=\frac{2^{13/4}}{\pi ^{1/2}}\left( K_1V\right) \epsilon
^{5/4}S^{-1/2}\left( 1+\frac{57}4\overline{K}_2-\frac 14\epsilon \right)
e^{-S_{cl}},  \eqnum{21}
\end{equation}
where the WKB exponent or the classical action $S_{cl}$ has been presented
in Eq. (17).

Now we apply the effective Hamiltonian approach to evaluate the ground-state
tunnel splitting.\cite{19} For the present case, the effective Hamiltonian
can be written as 
\begin{equation}
H_{eff}=\left[ 
\begin{array}{ll}
0 & -\hbar \Delta _0 \\ 
-\hbar \Delta _0 & 0
\end{array}
\right] .  \eqnum{22}
\end{equation}
A simple diagonalization of $H_{eff}$ shows that the eigenvalues of this
system are $\pm \hbar \Delta _0$. Therefore, the splitting of ground state
due to resonant coherently quantum tunneling of the magnetization vector
between energetically degenerate states is $\hbar \Delta =2\hbar \Delta _0$,
where $\hbar \Delta _0$ is shown in Eq. (21) with Eq. (17) for single-domain
FM particles with trigonal crystal symmetry in a magnetic field applied
perpendicular to the anisotropy axis $\left( \theta _H=\pi /2\right) $.

\subsection*{B. $\pi /2<\theta _H<\pi $}

For $\pi /2<\theta _H<\pi $, the critical angle $\theta _c$ is in the range
of $0<\theta _c<\pi /2$, and $\eta \approx \sqrt{2\epsilon /3}$. Then $%
\overline{E}_1\left( \delta \right) $ of Eq. (14) reduces to 
\begin{equation}
\overline{E}_1\left( \delta \right) =\frac 12\frac{\left| \cot \theta
_H\right| ^{1/3}}{1+\left| \cot \theta _H\right| ^{2/3}}\left[ 1-\frac{15}2%
\overline{K}_2\frac 1{\left( 1+\left| \cot \theta _H\right| ^{2/3}\right)
^{1/2}}\right] \left( \sqrt{6\epsilon }\delta ^2-\delta ^3\right) . 
\eqnum{23}
\end{equation}
The dependence of the effective potential $\overline{E}_1\left( \delta
\right) $ on $\delta \left( =\theta -\theta _0\right) $ for $\theta _H=3\pi
/4$ is plotted in Fig. 2. Here, $\overline{K}_2=0.001$. Now the problem
becomes one of MQT, where the magnetization vector escapes from the
metastable state at $\delta =0$, $\phi =0$ through the barrier by quantum
tunneling. Substituting Eq. (23) into the classical equations of motion, the
classical solution called bounce is found to be 
\begin{eqnarray}
\overline{\phi } &=&i\left( 6\epsilon \right) ^{3/4}\left| \cot \theta
_H\right| ^{1/6}\left( 1+\left| \cot \theta _H\right| ^{2/3}\right)
^{1/2}\left[ 1+\frac \epsilon 2-\frac 92\overline{K}_2\left( 1+\left| \cot
\theta _H\right| ^{2/3}\right) ^{1/2}\right.   \nonumber \\
&&\left. +\frac{\overline{K}_2}4\frac{2\left| \cot \theta _H\right| ^{2/3}-9%
}{\left( 1+\left| \cot \theta _H\right| ^{2/3}\right) ^{1/2}}+\overline{K}_2%
\frac{\left| \cot \theta _H\right| ^{2/3}-3}{\left( 1+\left| \cot \theta
_H\right| ^{2/3}\right) ^{3/2}}\right] \frac{\sinh \left( \overline{\omega }%
_c\overline{\tau }\right) }{\cosh ^3\left( \overline{\omega }_c\overline{%
\tau }\right) },  \nonumber \\
\overline{\delta } &=&\sqrt{6\epsilon }/\cosh ^2\left( \overline{\omega }_c%
\overline{\tau }\right) ,  \eqnum{24}
\end{eqnarray}
which corresponds to the variation of $\delta $ from $\delta =0$ at $\tau
=-\infty $ to the turning point $\delta =\sqrt{6\epsilon }$ at $\tau =0$,
and then back to $\delta =0$ at $\tau =+\infty $, where 
\begin{eqnarray}
\overline{\omega }_c &=&3^{1/4}\times 2^{-3/4}\epsilon ^{1/4}\frac{\left|
\cot \theta _H\right| ^{1/6}}{1+\left| \cot \theta _H\right| ^{2/3}}\left[ 1-%
\frac \epsilon 2+\frac 92\overline{K}_2\left( 1+\left| \cot \theta _H\right|
^{2/3}\right) ^{1/2}\right.   \nonumber \\
&&\left. +\frac{\overline{K}_2}4\frac{2\left| \cot \theta _H\right| ^{2/3}-21%
}{\left( 1+\left| \cot \theta _H\right| ^{2/3}\right) ^{1/2}}+\overline{K}_2%
\frac{\left| \cot \theta _H\right| ^{2/3}+3}{\left( 1+\left| \cot \theta
_H\right| ^{2/3}\right) ^{3/2}}\right] .  \nonumber
\end{eqnarray}
The associated classical action is then given by 
\begin{eqnarray}
S_{cl} &=&\frac{3^{1/4}\times 2^{17/4}}5S\epsilon ^{5/4}\left| \cot \theta
_H\right| ^{1/6}\left[ 1+\frac \epsilon 2-\frac 92\overline{K}_2\left(
1+\left| \cot \theta _H\right| ^{2/3}\right) ^{1/2}\right.   \nonumber \\
&&\left. -\frac{\overline{K}_2}2\frac{\left| \cot \theta _H\right| ^{2/3}+9/2%
}{\left( 1+\left| \cot \theta _H\right| ^{2/3}\right) ^{1/2}}-\overline{K}_2%
\frac{\left| \cot \theta _H\right| ^{2/3}+3}{\left( 1+\left| \cot \theta
_H\right| ^{2/3}\right) ^{3/2}}\right] .  \eqnum{25}
\end{eqnarray}
For this case, the barrier height is 
\begin{eqnarray*}
U &=&2K_1V\overline{E}_1\left( \delta _m\right)  \\
&=&\frac{2^{7/2}}{3^{3/2}}\frac{\left| \cot \theta _H\right| ^{1/3}}{%
1+\left| \cot \theta _H\right| ^{2/3}}\left[ 1-\frac{15}2\overline{K}_2\frac %
1{\left( 1+\left| \cot \theta _H\right| ^{2/3}\right) ^{1/2}}\right]
\epsilon ^{3/2}\left( K_1V\right) ,
\end{eqnarray*}
at $\delta _m=2\left( 6\epsilon \right) ^{1/2}/3$, and the frequency of
small oscillations of the magnetization vector around the bottom of the
metastable well is 
\begin{eqnarray*}
\overline{\omega }_b &=&3^{1/4}\times 2^{1/4}\epsilon ^{1/4}\frac{\left|
\cot \theta _H\right| ^{1/6}}{1+\left| \cot \theta _H\right| ^{2/3}}\left[ 1-%
\frac \epsilon 2+\frac 92\overline{K}_2\left( 1+\left| \cot \theta _H\right|
^{2/3}\right) ^{1/2}\right.  \\
&&\left. +\frac{\overline{K}_2}4\frac{2\left| \cot \theta _H\right| ^{2/3}-21%
}{\left( 1+\left| \cot \theta _H\right| ^{2/3}\right) ^{1/2}}+\overline{K}_2%
\frac{\left| \cot \theta _H\right| ^{2/3}+3}{\left( 1+\left| \cot \theta
_H\right| ^{2/3}\right) ^{3/2}}\right]  \\
&=&2\overline{\omega }_c.
\end{eqnarray*}
Then the WKB exponent or the classical action $B$ is approximately given by 
\begin{eqnarray}
B &\backsim &\frac U{\hbar \omega _b}  \nonumber \\
&=&\frac{2^{9/4}}{3^{7/4}}S\epsilon ^{5/4}\left| \cot \theta _H\right|
^{1/6}\left[ 1+\frac \epsilon 2-\frac 92\overline{K}_2\left( 1+\left| \cot
\theta _H\right| ^{2/3}\right) ^{1/2}\right.   \nonumber \\
&&\left. -\frac{\overline{K}_2}2\frac{\left| \cot \theta _H\right| ^{2/3}+9/2%
}{\left( 1+\left| \cot \theta _H\right| ^{2/3}\right) ^{1/2}}-\overline{K}_2%
\frac{\left| \cot \theta _H\right| ^{2/3}+3}{\left( 1+\left| \cot \theta
_H\right| ^{2/3}\right) ^{3/2}}\right] ,  \eqnum{26}
\end{eqnarray}
which is consistent with Eq. (25) up to the numerical factor. After a simple
calculation, we obtain the crossover temperature as 
\begin{eqnarray}
k_BT_c &=&\frac 5{2^{3/4}\times 3^{7/4}}\epsilon ^{1/4}\frac{K_1V}S\frac{%
\left| \cot \theta _H\right| ^{1/6}}{1+\left| \cot \theta _H\right| ^{2/3}}%
\left[ 1-\frac \epsilon 2+\frac 92\overline{K}_2\left( 1+\left| \cot \theta
_H\right| ^{2/3}\right) ^{1/2}\right.   \nonumber \\
&&\left. +\frac{\overline{K}_2}2\frac{\left| \cot \theta _H\right|
^{2/3}-21/2}{\left( 1+\left| \cot \theta _H\right| ^{2/3}\right) ^{1/2}}+%
\overline{K}_2\frac{\left| \cot \theta _H\right| ^{2/3}+3}{\left( 1+\left|
\cot \theta _H\right| ^{2/3}\right) ^{3/2}}\right] ,  \eqnum{27}
\end{eqnarray}
corresponding to the transition from quantum to thermal regime. For a
nanometer-scale single-domain FM particle, the typical values of parameters
for the magnetic anisotropy coefficients are $K_1=10^8$ erg/cm$^3$, and $%
K_2=10^5$ erg/cm$^3$. The radius of the FM particle is about 12 nm and the
sublattice spin is $10^6$. If $\epsilon =0.001$, we obtain that $T_c\left(
135^{\circ }\right) \backsim 203$mK corresponding to the crossover from
quantum to classical regime, which compares well with the experimental
result of $0.31$K on single-domain FM nanoparticles of Barium ferrite
(BaFeCoTiO).\cite{7} Note that, even for $\epsilon $ as small as $10^{-3}$,
the angle corresponding to an appreciable change of the orientation of the
magnetization vector by quantum tunneling is $\delta _2=\sqrt{6\epsilon }$
rad$>4^{\circ }$.

The classical action $S_{cl}$ can be obtained by solving numerically the
equations of motion (4a) and (4b). In Fig. 3 we present the $\theta _H$
dependence of $S_{cl}$ with $\epsilon =0.001$ and $\overline{K}_2=0.001$ for 
$\pi /2<\theta _H<\pi $ by numerical and analytical calculations,
respectively. As is noted in the figure, the analytical result obtained from
Eq. (25) is almost valid in the whole range of angles $\pi /2<\theta _H<\pi $%
.

By applying the formulas in Ref. 12, and using Eq. (25) for the WKB exponent
or the classical action, we obtain the tunneling rate $\Gamma $ of the
magnetization vector in single-domain FM nanoparticles with trigonal crystal
symmetry in a magnetic field applied in the range of $\pi /2<\theta _H<\pi $
as 
\begin{eqnarray}
\Gamma &=&\frac{2^{31/8}\times 3^{7/8}}{\pi ^{1/2}}\frac V\hbar
K_1S^{-1/2}\epsilon ^{7/8}\frac{\left| \cot \theta _H\right| ^{1/4}}{%
1+\left| \cot \theta _H\right| ^{2/3}}\left[ 1-\frac \epsilon 4+\frac 94%
\overline{K}_2\left( 1+\left| \cot \theta _H\right| ^{2/3}\right)
^{1/2}\right.  \nonumber \\
&&\left. +\frac{\overline{K}_2}4\frac{\left| \cot \theta _H\right|
^{2/3}-51/2}{\left( 1+\left| \cot \theta _H\right| ^{2/3}\right) ^{1/2}}+%
\frac{\overline{K}_2}2\frac{\left| \cot \theta _H\right| ^{2/3}+3}{\left(
1+\left| \cot \theta _H\right| ^{2/3}\right) ^{3/2}}\right] e^{-S_{cl}}. 
\eqnum{28}
\end{eqnarray}

\section*{IV. MQC and MQT for hexagonal crystal symmetry}

In this section, we study the quantum tunneling of the magnetization vector
in single-domain FM particles with hexagonal crystal symmetry whose
magnetocrystalline anisotropy energy $E_a\left( \theta ,\phi \right) $ at
zero magnetic field can be written as 
\begin{equation}
E_a\left( \theta ,\phi \right) =K_1\sin ^2\theta +K_2\sin ^4\theta +K_3\sin
^6\theta -K_3^{\prime }\sin ^6\theta \cos \left( 6\phi \right) ,  \eqnum{29}
\end{equation}
where $K_1$, $K_2$, $K_3$, and $K_3^{\prime }$ are the magnetic anisotropic
coefficients. The easy axes are $\pm \widehat{z}$ for $K_1>0$. When we apply
an external magnetic field at an arbitrarily directed angle in the $ZX$
plane, the total energy of this system is given by 
\begin{equation}
E\left( \theta ,\phi \right) =E_a\left( \theta ,\phi \right) -M_0H_x\sin
\theta \cos \phi -M_0H_z\cos \theta +E_0,  \eqnum{30}
\end{equation}
By choosing $K_3^{\prime }>0$, we take $\phi =0$ to be the easy plane, at
which the potential energy can be written in terms of the dimensionless
parameters as 
\begin{equation}
\overline{E}\left( \theta ,\phi =0\right) =\frac 12\sin ^2\theta +\overline{K%
}_2\sin ^4\theta +\left( \overline{K}_3-\overline{K}_3^{\prime }\right) \sin
^6\theta -\overline{H}\cos \left( \theta -\theta _H\right) +\overline{E}_0, 
\eqnum{31}
\end{equation}
where $\overline{K}_3=K_3/2K_1$ and $\overline{K}_3^{\prime }=K_3^{\prime
}/2K_1$.

Then the initial angle $\theta _0$ is determined by $\left[ d\overline{E}%
\left( \theta ,0\right) /d\theta \right] _{\theta =\theta _0}=0$, and the
critical angle $\theta _c$ and the dimensionless critical field $\overline{H}%
_c$ by both $\left[ d\overline{E}\left( \theta ,0\right) /d\theta \right]
_{\theta =\theta _c,\overline{H}=\overline{H}_c}=0$ and $\left[ d^2\overline{%
E}\left( \theta ,0\right) /d\theta ^2\right] _{\theta =\theta _c,\overline{H}%
=\overline{H}_c}=0$, which leads to 
\begin{eqnarray}
&&\left. \frac 12\sin \left( 2\theta _0\right) +\overline{H}\sin \left(
\theta _0-\theta _H\right) +4\overline{K}_2\sin ^4\theta _0+6\left( 
\overline{K}_3-\overline{K}_3^{\prime }\right) \sin ^5\theta _0\cos \theta
_0=0,\right.  \eqnum{32a} \\
&&\left. \frac 12\sin \left( 2\theta _c\right) +\overline{H}_c\sin \left(
\theta _c-\theta _H\right) +4\overline{K}_2\sin ^4\theta _c+6\left( 
\overline{K}_3-\overline{K}_3^{\prime }\right) \sin ^5\theta _c\cos \theta
_c=0,\right.  \eqnum{32b} \\
&&\left. \cos \left( 2\theta _c\right) +\overline{H}_c\cos \left( \theta
_c-\theta _H\right) +4\overline{K}_2\left( 3\sin ^2\theta _c\cos ^2\theta
_c-\sin ^4\theta _c\right) \right.  \nonumber \\
&&\left. +6\left( \overline{K}_3-\overline{K}_3^{\prime }\right) \left(
5\sin ^4\theta _c\cos ^2\theta _c-\sin ^6\theta _c\right) =0,\right. 
\eqnum{32c}
\end{eqnarray}
Under the assumption that $\left| \overline{K}_2\right| $, $\left| \overline{%
K}_3-\overline{K}_3^{\prime }\right| \ll 1$, we obtain the dimensionless
critical field $\overline{H}_c$ as 
\begin{equation}
\overline{H}_c=\frac 1{\left[ \left( \sin \theta _H\right) ^{2/3}+\left|
\cos \theta _H\right| ^{2/3}\right] ^{3/2}}\left[ 1+\frac{4\overline{K}_2}{%
1+\left| \cot \theta _H\right| ^{2/3}}+\frac{6\left( \overline{K}_3-%
\overline{K}_3^{\prime }\right) }{\left( 1+\left| \cot \theta _H\right|
^{2/3}\right) ^2}\right] .  \eqnum{33}
\end{equation}
In the limit of small $\epsilon =1-\overline{H}/\overline{H}_c$, Eq. (32a)
becomes 
\begin{eqnarray}
&&-\epsilon \overline{H}_c\sin \left( \theta _c-\theta _H\right) +\eta
^2\left[ \left( 3/2\right) \overline{H}_c\sin \left( \theta _c-\theta
_H\right) +3\overline{K}_2\sin \left( 4\theta _c\right) \right.  \nonumber \\
&&\left. +12\left( \overline{K}_3-\overline{K}_3^{\prime }\right) \sin
^3\theta _c\cos \theta _c\left( 5-8\sin ^2\theta _c\right) \right] +\eta
\left\{ \epsilon \overline{H}_c\cos \left( \theta _c-\theta _H\right) \right.
\nonumber \\
&&-\eta ^2\left. \left[ \left( 1/2\right) \overline{H}_c\cos \left( \theta
_c-\theta _H\right) +4\overline{K}_2\cos \left( 4\theta _c\right) \right.
\right.  \nonumber \\
&&\left. \left. \left. +12\left( \overline{K}_3-\overline{K}_3^{\prime
}\right) \sin ^2\theta _c\left( 5-20\sin ^2\theta _c+16\sin ^4\theta
_c\right) \right] \right\} =0,\right.  \eqnum{34}
\end{eqnarray}
where $\eta \equiv \theta _c-\theta _0$ which is small for $\epsilon \ll 1$.
By introducing a small variable $\delta \equiv \theta -\theta _0$ $\left(
\left| \delta \right| \ll 1\text{ in the limit of }\epsilon \ll 1\right) $,
the total energy becomes 
\begin{equation}
\overline{E}\left( \delta ,\phi \right) =\overline{K}_3^{\prime }\left[
1-\cos \left( 6\phi \right) \right] \sin ^6\left( \theta _0+\delta \right) +%
\overline{H}_x\left( 1-\cos \phi \right) \sin \left( \theta _0+\delta
\right) +\overline{E}_1\left( \delta \right) ,  \eqnum{35}
\end{equation}
where $\overline{E}_1\left( \delta \right) $ is a function of only $\delta $
given by 
\begin{eqnarray}
\overline{E}_1\left( \delta \right) &=&\left[ \frac 12\overline{H}_c\sin
\left( \theta _c-\theta _H\right) +\overline{K}_2\sin \left( 4\theta
_c\right) +4\left( \overline{K}_3-\overline{K}_3^{\prime }\right) \left(
5\sin ^3\theta _c\cos ^3\theta _c-3\sin ^5\theta _c\cos \theta _c\right)
\right]  \nonumber \\
&&\times \left( \delta ^3-3\delta ^2\eta \right) +\left[ \frac 18\overline{H}%
_c\cos \left( \theta _c-\theta _H\right) +\overline{K}_2\cos \left( 4\theta
_c\right) +3\left( \overline{K}_3-\overline{K}_3^{\prime }\right) \sin
^2\theta _c\left( \sin ^4\theta _c\right. \right.  \nonumber \\
&&\left. \left. -10\sin ^2\theta _c\cos ^2\theta _c+5\cos ^4\theta _c\right)
\right] \left( \delta ^4-4\delta ^3\eta +6\delta ^2\eta ^2-4\delta
^2\epsilon \right) +\epsilon \delta ^2\left[ 4\overline{K}_2\cos \left(
4\theta _c\right) \right.  \nonumber \\
&&\left. +12\left( \overline{K}_3-\overline{K}_3^{\prime }\right) \sin
^2\theta _c\left( \sin ^4\theta _c-10\sin ^2\theta _c\cos ^2\theta _c+5\cos
^4\theta _c\right) \right] .  \eqnum{36}
\end{eqnarray}

In the following we investigate the MQC and MQT of the magnetization vector
in FM particles with hexagonal crystal symmetry for different angle ranges
of the external magnetic field: $\theta _H=\pi /2$, $\pi /2+O\left( \epsilon
^{3/2}\right) <\theta _H<\pi -O\left( \epsilon ^{3/2}\right) $, and $\theta
_H=\pi $, respectively.

\subsection*{A. $\theta _H=\pi /2$}

For $\theta _H=\pi /2$, i.e., the external magnetic field is applied
perpendicular to the anisotropy axis, we obtain that $\theta _c=\pi /2$ and $%
\eta =\sqrt{2\epsilon }\left[ 1-4\overline{K}_2-12\left( \overline{K}_3-%
\overline{K}_3^{\prime }\right) \right] $. Now $\overline{E}_1\left( \delta
\right) $ becomes 
\begin{equation}
\overline{E}_1\left( \delta \right) =\frac 18\left[ 1+12\overline{K}%
_2+30\left( \overline{K}_3-\overline{K}_3^{\prime }\right) \right] \delta
^2\left\{ \delta -2\sqrt{2\epsilon }\left[ 1-4\overline{K}_2-12\left( 
\overline{K}_3-\overline{K}_3^{\prime }\right) \right] \right\} ^2. 
\eqnum{37}
\end{equation}
Substituting Eq. (37) into the classical equations of motion, we obtain the
following instanton solution 
\begin{eqnarray}
\overline{\phi } &=&i\epsilon \left[ 1+\frac \epsilon 2-4\overline{K}_2-18%
\overline{K}_3^{\prime }-6\left( \overline{K}_3-\overline{K}_3^{\prime
}\right) \right] \frac 1{\cosh ^2\left( \overline{\omega }_c\overline{\tau }%
\right) },  \nonumber \\
\overline{\delta } &=&\sqrt{2\epsilon }\left[ 1-4\overline{K}_2-12\left( 
\overline{K}_3-\overline{K}_3^{\prime }\right) \right] \left[ 1+\tanh \left( 
\overline{\omega }_c\overline{\tau }\right) \right] ,  \eqnum{38}
\end{eqnarray}
which corresponds to the variation of $\delta $ from $\delta =0$ at $\tau
=-\infty $ to $\delta =2\sqrt{2\epsilon }\left[ 1-4\overline{K}_2-12\left( 
\overline{K}_3-\overline{K}_3^{\prime }\right) \right] $ at $\tau =+\infty $%
, where 
\[
\overline{\omega }_c=\sqrt{\frac \epsilon 2}\left[ 1-\frac \epsilon 2+4%
\overline{K}_2+18\overline{K}_3^{\prime }+6\left( \overline{K}_3-\overline{K}%
_3^{\prime }\right) \right] . 
\]
We can calculate the classical action by integrating the Euclidean action of
Eq. (2) with the above instanton solution, and the result is found to be 
\begin{equation}
S_{cl}=\frac{2^{5/2}}3S\epsilon ^{3/2}\left[ 1+\frac \epsilon 2-8\overline{K}%
_2-18\overline{K}_3^{\prime }-24\left( \overline{K}_3-\overline{K}_3^{\prime
}\right) \right] .  \eqnum{39}
\end{equation}
From Eq. (37) we obtain that the height of barrier is $U=2K_1V\overline{E}%
_1\left( \delta _m\right) =K_1V\epsilon ^2\left[ 1-4\overline{K}_2-18\left( 
\overline{K}_3-\overline{K}_3^{\prime }\right) \right] $ at $\delta _m=\sqrt{%
2\epsilon }\left[ 1-4\overline{K}_2-12\left( \overline{K}_3-\overline{K}%
_3^{\prime }\right) \right] $, and the oscillation frequency around the
minimum of the inverted potential $-\overline{E}_1\left( \delta \right) $ is 
\[
\overline{\omega }_b=\sqrt{\epsilon }\left[ 1-\frac \epsilon 2+4\overline{K}%
_2+18\overline{K}_3^{\prime }+6\left( \overline{K}_3-\overline{K}_3^{\prime
}\right) \right] =\sqrt{2}\overline{\omega }_c. 
\]
Then the WKB exponent is approximately given by 
\begin{eqnarray}
B &\backsim &\frac U{\hbar \omega _b}  \nonumber \\
&=&\frac 12S\epsilon ^{3/2}\left[ 1+\frac \epsilon 2-8\overline{K}_2-18%
\overline{K}_3^{\prime }-24\left( \overline{K}_3-\overline{K}_3^{\prime
}\right) \right] ,  \eqnum{40}
\end{eqnarray}
which agrees up to the numerical factor with Eq. (39) obtained by applying
the explicit instanton solution. The temperature corresponding to the
crossover from the quantum coherence between the degenerate ground-state
levels (coherent tunneling) to the classical over-barrier transition
(incoherent transition) is found to be 
\begin{equation}
k_BT_c=\frac 3{2^{5/2}}\epsilon ^{1/2}\frac{K_1V}S\left[ 1-\frac \epsilon 2+4%
\overline{K}_2+18\overline{K}_3^{\prime }+6\left( \overline{K}_3-\overline{K}%
_3^{\prime }\right) \right] ,  \eqnum{41}
\end{equation}
and the temperature corresponding to the crossover from quantum coherence
between the degenerate ground-state levels (coherent tunneling) to quantum
tunneling from excited levels (thermally assisted tunneling or incoherent
tunneling) is found to be 
\begin{equation}
k_BT_c^{\prime }=\frac 1\pi \epsilon ^{1/2}\frac{K_1V}S\left[ 1-\frac %
\epsilon 2+4\overline{K}_2+18\overline{K}_3^{\prime }+6\left( \overline{K}_3-%
\overline{K}_3^{\prime }\right) \right] .  \eqnum{42}
\end{equation}
By applying the instanton technique for single-domain FM particles in the
spin-coherent-state path-integral representation,\cite{12} we obtain the
instanton's contribution to the tunnel splitting, $\hbar \Delta _0$ as 
\begin{equation}
\hbar \Delta _0=\frac{2^{13/4}}{\pi ^{1/2}}\left( VK_1\right)
S^{-1/2}\epsilon ^{5/4}\left[ 1-\frac \epsilon 4+9\overline{K}_3^{\prime
}-6\left( \overline{K}_3-\overline{K}_3^{\prime }\right) \right] e^{-S_{cl}},
\eqnum{43}
\end{equation}
where the WKB exponent or the classical action $S_{cl}$ is clearly shown in
Eq. (39). Then the splitting of ground state due to resonant coherently
quantum tunneling of the magnetization vector between energetically
degenerate states is found to be $\hbar \Delta =2\hbar \Delta _0$ for FM
particles with hexagonal crystal symmetry in a magnetic field applied
perpendicular to the anisotropy axis $\left( \theta _H=\pi /2\right) $ with
the help of the effective Hamiltonian approach.

\subsection*{B. $\pi /2+O\left( \epsilon ^{3/2}\right) <\theta _H<\pi
-O\left( \epsilon ^{3/2}\right) $}

For this case, $\eta \approx \sqrt{2\epsilon }/3$ and the critical angle $%
\theta _c$ is found to be 
\[
\sin \theta _c=\frac 1{\left( 1+\left| \cot \theta _H\right| ^{2/3}\right)
^{1/2}}\left[ 1+\frac 83\overline{K}_2\frac{\left| \cot \theta _H\right|
^{2/3}}{1+\left| \cot \theta _H\right| ^{2/3}}+8\left( \overline{K}_3-%
\overline{K}_3^{\prime }\right) \frac{\left| \cot \theta _H\right| ^{2/3}}{%
\left( 1+\left| \cot \theta _H\right| ^{2/3}\right) ^2}\right] . 
\]
Now $\overline{E}_1\left( \delta \right) $ becomes 
\begin{eqnarray}
\overline{E}_1\left( \delta \right) &=&\frac 12\frac{\left| \cot \theta
_H\right| ^{1/3}}{1+\left| \cot \theta _H\right| ^{2/3}}\left[ 1-\frac 43%
\overline{K}_2\frac{7-4\left| \cot \theta _H\right| ^{2/3}}{1+\left| \cot
\theta _H\right| ^{2/3}}\right.  \nonumber \\
&&\left. +2\left( \overline{K}_3-\overline{K}_3^{\prime }\right) \frac{%
11-16\left| \cot \theta _H\right| ^{2/3}}{\left( 1+\left| \cot \theta
_H\right| ^{2/3}\right) ^2}\right] \left( \sqrt{6\epsilon }\delta ^2-\delta
^3\right) .  \eqnum{44}
\end{eqnarray}
Then the classical equations of motion have the following bounce solution 
\begin{eqnarray}
\overline{\phi } &=&i\left( 6\epsilon \right) ^{3/4}\left| \cot \theta
_H\right| ^{1/6}\left( 1+\left| \cot \theta _H\right| ^{2/3}\right)
^{1/2}\left[ 1+\frac \epsilon 2-\frac 43\overline{K}_2\frac{5-\left| \cot
\theta _H\right| ^{2/3}}{1+\left| \cot \theta _H\right| ^{2/3}}\right. 
\nonumber \\
&&\left. -18\overline{K}_3^{\prime }\frac 1{1+\left| \cot \theta _H\right|
^{2/3}}-2\left( \overline{K}_3-\overline{K}_3^{\prime }\right) \frac{%
7-6\left| \cot \theta _H\right| ^{2/3}}{\left( 1+\left| \cot \theta
_H\right| ^{2/3}\right) ^2}\right] \frac{\sinh \left( \overline{\omega }_c%
\overline{\tau }\right) }{\cosh ^3\left( \overline{\omega }_c\overline{\tau }%
\right) },  \nonumber \\
\overline{\delta } &=&\sqrt{6\epsilon }/\cosh ^2\left( \overline{\omega }_c%
\overline{\tau }\right) ,  \eqnum{45}
\end{eqnarray}
corresponding to the variation of $\delta $ from $\delta =0$ at $\tau
=-\infty $ to the turning point $\delta =\sqrt{6\epsilon }$ at $\tau =0$,
and then back to $\delta =0$ at $\tau =+\infty $, where 
\begin{eqnarray*}
\overline{\omega }_c &=&\left( \frac 38\right) ^{1/4}\epsilon ^{1/4}\frac{%
\left| \cot \theta _H\right| ^{1/6}}{1+\left| \cot \theta _H\right| ^{2/3}}%
\left[ 1-\frac \epsilon 2+\frac 43\overline{K}_2\frac{5-3\left| \cot \theta
_H\right| ^{2/3}}{1+\left| \cot \theta _H\right| ^{2/3}}\right. \\
&&\left. +18\overline{K}_3^{\prime }\frac 1{1+\left| \cot \theta _H\right|
^{2/3}}+2\left( \overline{K}_3-\overline{K}_3^{\prime }\right) \frac{%
7-10\left| \cot \theta _H\right| ^{2/3}}{\left( 1+\left| \cot \theta
_H\right| ^{2/3}\right) ^2}\right] .
\end{eqnarray*}
The classical action associated with this bounce solution is found to be 
\begin{eqnarray}
S_{cl} &=&\frac{2^{17/4}\times 3^{1/4}}5S\epsilon ^{5/4}\left| \cot \theta
_H\right| ^{1/6}\left[ 1+\frac \epsilon 2+\frac 43\overline{K}_2\frac{%
2-\left| \cot \theta _H\right| ^{2/3}}{1+\left| \cot \theta _H\right| ^{2/3}}%
\right.  \nonumber \\
&&\left. -18\overline{K}_3^{\prime }\frac 1{1+\left| \cot \theta _H\right|
^{2/3}}+4\left( \overline{K}_3-\overline{K}_3^{\prime }\right) \frac{%
2-3\left| \cot \theta _H\right| ^{2/3}}{\left( 1+\left| \cot \theta
_H\right| ^{2/3}\right) ^2}\right] .  \eqnum{46}
\end{eqnarray}

For this case, the barrier height $U\left( =2K_1V\overline{E}_1\left( \delta
_m=2\sqrt{6\epsilon }/3\right) \right) $ is given by 
\begin{eqnarray*}
U &=&\frac{2^{7/2}}{3^{3/2}}\left( K_1V\right) \epsilon ^{3/2}\frac{\left|
\cot \theta _H\right| ^{1/3}}{1+\left| \cot \theta _H\right| ^{2/3}}\left[ 1+%
\frac 43\overline{K}_2\frac{7-4\left| \cot \theta _H\right| ^{2/3}}{1+\left|
\cot \theta _H\right| ^{2/3}}\right. \\
&&\left. +2\left( \overline{K}_3-\overline{K}_3^{\prime }\right) \frac{%
11-16\left| \cot \theta _H\right| ^{2/3}}{\left( 1+\left| \cot \theta
_H\right| ^{2/3}\right) ^2}\right] ,
\end{eqnarray*}
and the frequency of small oscillations of the magnetization vector around
the minimum of the inverted potential $-\overline{E}_1\left( \delta \right) $
is 
\begin{eqnarray*}
\overline{\omega }_b &=&3^{1/4}\times 2^{1/4}\epsilon ^{1/4}\frac{\left|
\cot \theta _H\right| ^{1/6}}{1+\left| \cot \theta _H\right| ^{2/3}}\left[ 1-%
\frac \epsilon 2+\frac 43\overline{K}_2\frac{5-3\left| \cot \theta _H\right|
^{2/3}}{1+\left| \cot \theta _H\right| ^{2/3}}\right. \\
&&\left. +18\overline{K}_3^{\prime }\frac 1{1+\left| \cot \theta _H\right|
^{2/3}}+2\left( \overline{K}_3-\overline{K}_3^{\prime }\right) \frac{%
7-10\left| \cot \theta _H\right| ^{2/3}}{\left( 1+\left| \cot \theta
_H\right| ^{2/3}\right) ^2}\right] \\
&=&2\overline{\omega }_c.
\end{eqnarray*}
Then the WKB\ exponent is approximately given by 
\begin{eqnarray}
B &\backsim &\frac U{\hbar \omega _b}  \nonumber \\
&=&\frac{2^{9/4}}{3^{7/4}}S\epsilon ^{5/4}\left| \cot \theta _H\right|
^{1/6}\left[ 1+\frac \epsilon 2+\frac 43\overline{K}_2\frac{2-\left| \cot
\theta _H\right| ^{2/3}}{1+\left| \cot \theta _H\right| ^{2/3}}\right. 
\nonumber \\
&&\left. -18\overline{K}_3^{\prime }\frac 1{1+\left| \cot \theta _H\right|
^{2/3}}+4\left( \overline{K}_3-\overline{K}_3^{\prime }\right) \frac{%
2-3\left| \cot \theta _H\right| ^{2/3}}{\left( 1+\left| \cot \theta
_H\right| ^{2/3}\right) ^2}\right] ,  \eqnum{47}
\end{eqnarray}
which agrees with Eq. (46) up to the numerical factor. Equating the
classical action $S_{cl}$ to $U/k_BT_c$, where $U$ is the barrier height, we
obtain that the crossover from quantum to classical behavior occurs at 
\begin{eqnarray}
k_BT_c &=&\frac 5{2^{3/4}\times 3^{7/4}}\epsilon ^{1/4}\frac{K_1V}S\frac{%
\left| \cot \theta _H\right| ^{1/6}}{1+\left| \cot \theta _H\right| ^{2/3}}%
\left[ 1-\frac \epsilon 2+\frac 43\overline{K}_2\frac{5-3\left| \cot \theta
_H\right| ^{2/3}}{1+\left| \cot \theta _H\right| ^{2/3}}\right.  \nonumber \\
&&\left. +18\overline{K}_3^{\prime }\frac 1{1+\left| \cot \theta _H\right|
^{2/3}}+2\left( \overline{K}_3-\overline{K}_3^{\prime }\right) \frac{%
7-10\left| \cot \theta _H\right| ^{2/3}}{\left( 1+\left| \cot \theta
_H\right| ^{2/3}\right) ^2}\right] .  \eqnum{48}
\end{eqnarray}
Based on the instanton technique,\cite{12} we obtain the tunneling rate
corresponding to the escaping of the magnetization vector from the
metastable state for single-domain FM nanoparticles with hexagonal crystal
symmetry in a magnetic field applied in the range of $\pi /2+O\left(
\epsilon ^{3/2}\right) <\theta _H<\pi -O\left( \epsilon ^{3/2}\right) $ as
the following equation, 
\begin{eqnarray}
\Gamma &=&\frac{2^{31/8}\times 3^{7/8}}{\pi ^{1/2}}\frac V\hbar
K_1S^{-1/2}\epsilon ^{7/8}\frac{\left| \cot \theta _H\right| ^{1/4}}{%
1+\left| \cot \theta _H\right| ^{2/3}}\left[ 1-\frac \epsilon 4+9\overline{K}%
_3^{\prime }\frac 1{1+\left| \cot \theta _H\right| ^{2/3}}\right.  \nonumber
\\
&&\left. -\frac 23\overline{K}_2\frac{12-7\left| \cot \theta _H\right| ^{2/3}%
}{1+\left| \cot \theta _H\right| ^{2/3}}\right.  \nonumber \\
&&\left. +2\left( \overline{K}_3-\overline{K}_3^{\prime }\right) \frac{%
9-13\left| \cot \theta _H\right| ^{2/3}}{\left( 1+\left| \cot \theta
_H\right| ^{2/3}\right) ^2}\right] e^{-S_{cl}},  \eqnum{49}
\end{eqnarray}
where the WKB\ exponent or the classical action $S_{cl}$ has been clearly
shown in Eq. (46).

\subsection*{C. $\theta _H=\pi $}

Finally, we study the MQT of the magnetization vector corresponding to the
escaping from the metastable state in single-domain FM nanoparticles with
hexagonal crystal symmetry in a magnetic field applied at $\theta _H=\pi $,
i.e., antiparallel to the anisotropy axis. Now the total energy become 
\begin{eqnarray}
\overline{E}\left( \delta ,\phi \right) &=&\overline{K}_3^{\prime }\left[
1-\cos \left( 6\phi \right) \right] \delta ^6+\frac 12\delta ^2\left[
\epsilon -\frac 14\left( 1-8\overline{K}_2\right) \delta ^2\right]  \nonumber
\\
&&-\frac 1{24}\delta ^4\left\{ \epsilon -\frac 12\left[ 1-32\overline{K}%
_2+48\left( \overline{K}_3-\overline{K}_3^{\prime }\right) \right] \delta
^2\right\} .  \eqnum{50}
\end{eqnarray}
The classical equations of motion have the bounce solution 
\begin{eqnarray}
\overline{\phi } &=&-i\overline{\omega }_c\overline{\tau }+\frac{n\pi }3, 
\nonumber \\
\overline{\delta } &=&\sqrt{\frac{4\epsilon }{1-8\overline{K}_2-\epsilon
\left[ 32\overline{K}_3^{\prime }\left( 1-\cosh \left( 6\overline{\omega }_c%
\overline{\tau }\right) \right) +\frac 13\left( 1-48\overline{K}_2+96\left( 
\overline{K}_3-\overline{K}_3^{\prime }\right) \right) \right] }}, 
\eqnum{51}
\end{eqnarray}
where $n=0,1,2,3,4,5$, and $\overline{\omega }_c=\epsilon $. The
corresponding classical action is found to be 
\begin{equation}
S_{cl}=\frac 23S\epsilon \frac 1{\Delta _1}\ln \left( \frac{2\Delta _1}{%
\Delta _2}\right) ,  \eqnum{52}
\end{equation}
with 
\begin{equation}
\Delta _1=1-8\overline{K}_2-32\overline{K}_3^{\prime }\epsilon -\frac 13%
\epsilon \left[ 1-48\overline{K}_2+96\left( \overline{K}_3-\overline{K}%
_3^{\prime }\right) \right] ,  \eqnum{53}
\end{equation}
and 
\begin{equation}
\Delta _2=32\overline{K}_3^{\prime }\epsilon .  \eqnum{54}
\end{equation}
According to the formulas in Ref. 12, we obtain the tunneling rate of the
magnetization vector escaping from the metastable state for single-domain FM
nanoparticles with hexagonal crystal symmetry in a magnetic field applied
antiparallel to the anisotropy axis $\left( \theta _H=\pi \right) $ as 
\begin{eqnarray}
\Gamma &=&\frac{2^{13/2}\times 3^{1/2}}{\pi ^{1/2}}\frac V\hbar
K_1S^{-1/2}\epsilon \left( 1+4\overline{K}_2\right)  \nonumber \\
&&\frac 1{1-16\overline{K}_2-64\overline{K}_3^{\prime }\epsilon -\frac 23%
\left[ 1-48\overline{K}_2+96\left( \overline{K}_3-\overline{K}_3^{\prime
}\right) \right] }e^{-S_{cl}},  \eqnum{55}
\end{eqnarray}
where the WKB exponent or the classical action $S_{cl}$ is shown in Eq.
(52). Eq. (50) shows that in this case $\left| \phi \right| \ll 1$ is not
valid, and therefore the problem can not be reduced to the one-dimensional
motion problem. And the effective potential energy and the effective mass in
one-dimensional form are not appropriate for the present case.

Now we discuss the range of angles that Eq. (46) is valid. Introducing $%
\theta _1=\theta _H-\pi /2$ and $\theta _2=\pi -\theta _H$, from Eqs. (39),
(46) and (52), we find that $\theta _1\approx \left( 5^6\times
2^{-21/2}\times 3^{-15/2}\right) \epsilon ^{3/2}$ and $\theta _2\approx
\left( 5^{-6}\times 2^{39/2}\times 3^{15/2}\right) \epsilon ^{3/2}$. This
means that Eq. (46) is almost valid in a wide range of angles $91^{\circ
}\leq \theta _H\leq 179^{\circ \text{ }}$ for $\epsilon =0.001$.

For the single-domain FM nanoparticle with hexagonal crystal symmetry in the
presence of an external magnetic field at arbitrarily directed angle, by
using Eqs. (39) and (43) for $\theta _H=\pi /2$, Eqs. (46) and (49) for $\pi
/2+O\left( \epsilon ^{3/2}\right) <\theta _H<\pi -O\left( \epsilon
^{3/2}\right) $, and Eqs. (52) and (55) for $\theta _H=\pi $, we obtain the
ground-state tunnel splitting for MQC and the tunneling rate for MQT of the
magnetization vector. Our results show that the tunnel splitting and the
tunneling rate depend on the orientation of the external magnetic field
distinctly. When $\theta _H=\pi /2$, the magnetic field is applied
perpendicular to the anisotropy axis, and when $\theta _H=\pi $, the field
is antiparallel to the anisotropy axis. It is found that even a very small
misalignment of the field with the above two orientations can completely
change the results of tunneling rates. Another interesting observation
concerns the dependence of the WKB exponent or the classical action with the
strength of the external magnetic field. In a wide range of angles, the $%
\epsilon \left( =1-\overline{H}/\overline{H}_c\right) $ dependence of the
WKB exponent $S_{cl}$ is given by $\epsilon ^{5/4}$, not $\epsilon ^{3/2}$
for $\theta _H=\pi /2$, and $\epsilon $ for $\theta _H=\pi $. Therefore,
both the orientation and the strength of the external magnetic field are the
control parameters for the experimental test for MQT and MQC of the
magnetization vector in single-domain FM nanoparticles.

\section*{V. Conclusions}

In summary we have investigated the tunneling behaviors of the magnetization
vector in single-domain FM nanoparticles in the presence of an external
magnetic field at arbitrarily directed angle. We consider the
magnetocrystalline anisotropy with the trigonal crystal symmetry and that
with the hexagonal crystal symmetry. By applying the instanton technique in
the spin-coherent-state path-integral representation, we obtain both the WKB
exponent and the preexponential factors in the tunnel splitting between
energetically degenerate states in MQC and the tunneling rate escaping from
a metastable state in MQT of the magnetization vector in the low barrier
limit for the external magnetic field perpendicular to the easy axis $\left(
\theta _H=\pi /2\right) $, for the field antiparallel to the initial easy
axis $\left( \theta _H=\pi \right) $, and for the field at an angle between
these two orientations $\left( \pi /2+O\left( \epsilon ^{3/2}\right) <\theta
_H<\pi -O\left( \epsilon ^{3/2}\right) \right) $. One important conclusion
is that the tunneling rate and the tunnel splitting depend on the
orientation of the external magnetic field distinctly. Another interesting
conclusion concerns the field strength dependence of the WKB\ exponent or
the classical action. We have found that in a wide range of angles, the $%
\epsilon \left( =1-\overline{H}/\overline{H}_c\right) $ dependence of the
WKB exponent or the classical action $S_{cl}$ is given by $\epsilon ^{5/4}$,
not $\epsilon ^{3/2}$ for $\theta _H=\pi /2$, and $\epsilon $ for $\theta
_H=\pi $. We have obtained the temperatures corresponding to the crossover
from quantum to thermal regime which are found to depend on the orientation
of the external magnetic field distinctly. As a result, we conclude that
both the orientation and the strength of the external magnetic field are the
controllable parameters for the experimental test of the phenomena of
macroscopic quantum tunneling and coherence of the magnetization vector in
single-domain FM nanoparticles with trigonal and hexagonal symmetries at a
temperature well bellow the crossover temperature. We have analyzed the
validity of the semiclassical approximation performed in the present work,
and have found that the semiclassical approximation should be already rather
good for the typical values of parameters for single-domain FM nanoparticles.

Recently, Wernsdorfer and co-workers performed the switching field
measurements on individual ferrimagnetic and insulating BaFeCoTiO
nanoparticles containing about $10^5$-$10^6$ spins at very low temperatures
(0.1-6K).\cite{7} They found that above 0.4K, the magnetization reversal of
these particles is unambiguously described by the N\'{e}el-Brown theory of
thermal activated rotation of the particle's moment over a well defined
anisotropy energy barrier. Below 0.4K, strong deviations from this model are
evidenced which are quantitatively in agreement with the predictions of the
MQT theory without dissipation.\cite{3} The BaFeCoTiO nanoparticles have a
strong uniaxial magnetocrystalline anisotropy.\cite{7} However, the
theoretical results presented here may be useful for checking the general
theory in a wide range of systems, with more general symmetries. The
experimental procedures on single-domain FM\ nanoparticles of Barium ferrite
with uniaxial symmetry\cite{7} may be applied to the systems with more
general symmetries. Note that the inverse of the WKB exponent $B^{-1}$ is
the magnetic viscosity $S$ at the quantum-tunneling-dominated regime $T\ll
T_c$ studied by magnetic relaxation measurements.\cite{1} Therefore, the
quantum tunneling of the magnetization should be checked at any $\theta _H$
by magnetic relaxation measurements. Over the past years a lot of
experimental and theoretical works were performed on the spin tunneling in
molecular Mn$_{12}$-Ac\cite{22} and Fe$_8$\cite{23} clusters having a
collective spin state $S=10$ (in this paper $S=10^6$). Further experiments
should focus on the level quantization of collective spin states of $S=10^2$-%
$10^4$. We hope that the theoretical results presented in this paper may
stimulate more experiments whose aim is observing macroscopic quantum
phenomena in nanometer-scale single-domain ferromagnets.

\section*{Acknowledgments}

R.L. would like to acknowledge Dr. Su-Peng Kou, Professor Zhan Xu, Professor
Jiu-Qing Liang and Professor Fu-Cho Pu for stimulating discussions.

\section*{Appendix A: Evaluation of the preexponential factors in WKB
tunneling rate}

In this appendix, we review briefly the procedure on how to calculate the
preexponential factors in the WKB rate of quantum tunneling of the
magnetization vector in single-domain FM particles, based on the instanton
technique in the spin-coherent-state path-integral representation.\cite{12}
The preexponential factors in tunneling rate (MQT) or the tunnel splitting
(MQC) are due to the quantum fluctuations about the classical path, which
can be evaluated by expanding the Euclidean action to second order in small
fluctuations. Then we apply this approach to obtain the instanton's
contribution to the ground-state tunnel splitting for resonant coherently
quantum tunneling of the magnetization vector in FM particles with trigonal
crystal symmetry in an external magnetic field applied perpendicular to the
anisotropy axis (considered in Sec. III) in detail.

In Ref. 12, Garg and Kim have studied the general formulas for evaluating
both the WKB exponent and the preexponential factors in the tunneling rate
or the tunnel splitting in the single-domain FM particles based on the
instanton technique in the spin-coherent-state path-integral representation,
without assuming a specific form of the magnetocrystalline anisotropy and
the external magnetic field. Here we explain briefly the basic idea of this
calculation. Such a calculation consists of two major steps. The first step
is to find the classical, or least-action path (instanton) from the
classical equations of motion, which gives the exponent or the classical
action in the WKB tunneling rate. Instantons in one-dimensional field theory
can be viewed as pseudoparticles with trajectories existing in the energy
barrier, and are therefore responsible for quantum tunneling. The second
step is to expand the Euclidean action to second order in the small
fluctuations about the classical path, and then evaluate the Van Vleck
determinant of resulting quadratic form.\cite{11,12} For single-domain FM
particles, writing $\theta \left( \tau \right) =\overline{\theta }\left(
\tau \right) +\theta _1\left( \tau \right) $ and $\phi \left( \tau \right) =%
\overline{\phi }\left( \tau \right) +\phi _1\left( \tau \right) $, where $%
\overline{\theta }$ and $\overline{\phi }$ denote the classical path, one
obtains the Euclidean action of Eq. (2) as $S_E\left[ \theta \left( \tau
\right) ,\phi \left( \tau \right) \right] \approx S_{cl}+\delta ^2S$ with $%
S_{cl}$ being the classical action or the WKB exponent and $\delta ^2S$
being a functional of small fluctuations $\theta _1$ and $\phi _1$,\cite{12} 
\begin{eqnarray}
\delta ^2S &=&-iS\int \frac d{d\tau }\left[ \sin \overline{\theta }\theta
_1\right] \phi _1d\tau +\frac i2S\int \cos \overline{\theta }\left( \frac{d%
\overline{\phi }}{d\tau }\right) \theta _1^2d\tau  \nonumber \\
&&+\frac{V_0}{2\hbar }\int \left( E_{\theta \theta }\theta _1^2+2E_{\theta
\phi }\theta _1\phi _1+E_{\phi \phi }\phi _1^2\right) d\tau ,  \eqnum{A1}
\end{eqnarray}
where $E_{\theta \theta }=\left( \partial ^2E/\partial \theta ^2\right)
_{\theta =\overline{\theta },\phi =\overline{\phi }}$, $E_{\theta \phi
}=\left( \partial ^2E/\partial \theta \partial \phi \right) _{\theta =%
\overline{\theta },\phi =\overline{\phi }}$, and $E_{\phi \phi }=\left(
\partial ^2E/\partial \phi ^2\right) _{\theta =\overline{\theta },\phi =%
\overline{\phi }}$. Under the condition that $E_{\phi \phi }>0$, the
Gaussian integration can be performed over $\phi _1$, and the remaining $%
\theta _1$ path integral can be casted into the standard form for a
one-dimensional motion problem. As usual there exists a zero-mode, $d%
\overline{\theta }/d\tau $, corresponding to a translation of the center of
the instanton, and a negative eigenvalue in the MQT problem.\cite{11,12}
This leads to the imaginary part of the energy, which corresponds to the
quantum escaping rate from the metastable state through the classically
impenetrable barrier to a stable one. The resonant tunnel splittings of the
ground state for the MQC problem can be evaluated by applying the similar
technique. What is need for the calculation of the tunneling rate (in MQT)
and the tunnel splitting (in MQC) is the asymptotic relation of the zero
mode, $d\overline{\theta }/d\tau $, for large $\tau $,\cite{11,12} 
\begin{equation}
d\overline{\theta }/d\tau \approx ae^{-\mu \zeta },\text{ as }\zeta
\rightarrow \infty .  \eqnum{A2}
\end{equation}
The new time variable $\zeta $ in Eq. (A2) is related to $\tau $ as 
\begin{equation}
d\zeta =d\tau /2A\left( \overline{\theta }\left( \tau \right) ,\overline{%
\phi }\left( \tau \right) \right) ,  \eqnum{A3}
\end{equation}
where 
\begin{equation}
A\left( \overline{\theta },\overline{\phi }\right) =\hbar S^2\sin ^2%
\overline{\theta }/2VE_{\phi \phi }.  \eqnum{A4}
\end{equation}
The partial derivatives are evaluated at the classical path. Then the
instanton's contribution to the tunneling rate for MQT or the tunnel
splitting for MQC\ of the magnetization vector in single-domain FM
nanoparticles (without the contribution of the topological Wess-Zumino, or
Berry phase term in the Euclidean action) is given by\cite{11,12}

\begin{equation}
\left| a\right| \left( \mu /\pi \right) ^{1/2}e^{-S_{cl}}.  \eqnum{A5}
\end{equation}
Therefore, all that is necessary is to differentiate the classical path
(instanton) to obtain $d\overline{\theta }/d\tau $, then convert from $\tau $
to the new time variable $\zeta $ according to Eqs. (A3) and (A4), and read
off $a$ and $\mu $ by comparison with Eq. (A2). If the condition $E_{\phi
\phi }>0$ is not satisfied, one can always perform the Gaussian integration
over $\theta _1$ and end up with a one-dimensional path integral over $\phi
_1$.

Now we apply this approach to the problem of resonant coherently quantum
tunneling of the magnetization vector between energetically degenerate easy
directions in single-domain FM nanoparticle with trigonal crystal symmetry
in an external magnetic field applied perpendicular to the anisotropy axis.
After some algebra, we find that 
\begin{equation}
E_{\phi \phi }\approx 2K_1\left( 1+12\overline{K}_2-\epsilon \right) , 
\eqnum{A6}
\end{equation}
which is positive. So we can perform the Gaussian integration over $\phi _1$
directly. The relation between $\tau $ and the new imaginary-time variable $%
\zeta $ for this MQC problem is found to be 
\begin{equation}
\tau =\frac{\hbar S^2}{2K_1V\left( 1+12\overline{K}_2-\epsilon \right) }%
\zeta .  \eqnum{A7}
\end{equation}
It is easy to differentiate the instanton solution to obtain 
\begin{equation}
\frac{d\overline{\delta }}{d\tau }=8\frac{K_1V}{\hbar S}\epsilon \left( 1+15%
\overline{K}_2-\frac \epsilon 2\right) \exp \left[ -\sqrt{2\epsilon }S\left(
1-\frac 32\overline{K}_2+\frac \epsilon 2\right) \zeta \right] ,  \eqnum{A8}
\end{equation}
as $\zeta \rightarrow \infty $. Thus, 
\begin{equation}
\left| a\right| =8\frac{K_1V}{\hbar S}\epsilon \left( 1+15\overline{K}_2-%
\frac \epsilon 2\right) ,  \eqnum{A9}
\end{equation}
and 
\begin{equation}
\mu =\sqrt{2\epsilon }S\left( 1-\frac 32\overline{K}_2+\frac \epsilon 2%
\right) .  \eqnum{A10}
\end{equation}
Substituting Eqs. (A9) and (A10) into the general formula (A5), and using
Eq. (17) for the classical action or the WKB exponent, we obtain the
instanton's contribution to the tunnel splitting $\hbar \Delta _0$ as
expressed in Eq. (21) for nanometer-scale single-domain ferromagnets with
trigonal crystal symmetry in the presence of an external magnetic field
applied perpendicular to the anisotropy axis.

The calculations of the tunnel splitting and the tunneling rate of the
magnetization vector for other MQT and MQC problems considered in the
present work can be performed by applying the similar techniques, and we
will not discuss them in any further.

Figure Captions:

Fig. 1 The $\delta \left( =\theta -\theta _0\right) $ dependence of the
effective potential $\overline{E}_1\left( \delta \right) $ for $\theta
_H=\pi /2$ (MQC).

Fig. 2 The $\delta \left( =\theta -\theta _0\right) $ dependence of the
effective potential $\overline{E}_1\left( \delta \right) $ for $\theta
_H=3\pi /4$ (MQT). Here, $\overline{K}_2=0.001$.

Fig. 3 The $\theta _H$ dependence of the relative classical action $%
S_{cl}\left( \theta _H\right) /S_{cl}\left( \theta _H=3\pi /4\right) $ in
the trigonal symmetry with $\epsilon =0.001$ and $\overline{K}_2=0.001$ by
numerical and analytical calculations.


\begin{references}
\bibitem{1}  For a review, see {\it Quantum Tunneling of Magnetization},
edited by L. Gunther and B. Barbara (Kluwer, Dordrecht, 1995); and E. M.
Chudnovsky and J. Tejada, {\it Macroscopic Quantum Tunneling of the Magnetic
Moment} (Cambridge University Press, 1997).

\bibitem{2}  O. B. Zaslavskii, Phys. Rev. B {\bf 42}, 992 (1990).

\bibitem{3}  M. -G. Miguel and E. M. Chudnovsky, Phys. Rev. B {\bf 54}, 388
(1996).

\bibitem{4}  G. -H. Kim and D. S. Hwang, Phys. Rev. B {\bf 55}, 8918 (1997).

\bibitem{5}  G. -H. Kim, Phys. Rev. B {\bf 57}, 10688 (1998).

\bibitem{6}  D. A. Garanin, X. M. Hidalgo, and E. M. Chudonovsky, Phys. Rev.
B {\bf 57}, 13639 (1998).

\bibitem{7}  W. Wernsdorfer, E. B. Orozco, K. Hasselbach, A. Benoit, D.
Mailly, O. Kubo, H. Nakano, and B. Barbara, Phys. Rev. Lett. {\bf 79}, 4014
(1997).

\bibitem{8}  Rong L\"{u}, Jia-Lin Zhu, Xiao-Bing Wang, and Lee Chang, Phys.
Rev. B {\bf 60}, 4101 (1999).

\bibitem{9}  D. Loss, D. P. DiVicenzo, and G. Grinstein, Phys. Rev. Lett. 
{\bf 69}, 3232 (1992).

\bibitem{10}  J. V. Delft and G. L. Henley, Phys. Rev. Lett. {\bf 69}, 3236
(1992).

\bibitem{11}  S. Coleman, {\it Aspects of Symmetry} (Cambridge University
Press, Cambridge, England, 1985), Chap.7.

\bibitem{12}  A. Garg and G. -H. Kim, J. Appl. Phys. {\bf 67}, 5669 (1990);
Phys. Rev. B {\bf 45}, 12921 (1992).

\bibitem{13}  P. H\"{a}mggi, P. Talkner, and M. Borkovec, Rev. Mod. Phys. 
{\bf 62}, 251 (1990).

\bibitem{14}  E. M. Chudnovsky and D. A. Garanin, Phys. Rev. Lett. {\bf 79},
4469 (1997).

\bibitem{15}  E. M. Chudnovsky, Phys. Lett. A {\bf 46}, 8011 (1992); D. A.
Garanin and E. M. Chudnovsky, Phys. Rev. B {\bf 56}, 11102 (1997).

\bibitem{16}  J. -Q. Liang, H. J. W. M\"{u}ller-Kirstein, D. K. Park, and F.
Zimmerschied, Phys. Rev. Lett. {\bf 81}, 216 (1998); S. Y. Lee, H. J. W.
M\"{u}ller-Kirstein, D. K. Park, and F. Zimmerschied, Phys. Rev. B {\bf 58},
5554 (1998); H. J. W. M\"{u}ller-Kirstein, D. K. Park, and J. M. S. Rana,
cond-mat/9902184.

\bibitem{17}  D. A. Gorokhov and G. Blatter, Phys. Rev. B {\bf 56}, 3130
(1997).

\bibitem{18}  C. S. Park, S. -K. Yoo, D. K. Park, and D. -H. Yoon,
cond-mat/9807344; cond-matt/9902039; C. S. Park, S. -K. Yoo, and D. -H.
Yoon, cond-mat/9909217.

\bibitem{19}  Xiao-Bing Wang and Fu-Cho Pu, J. Phys.: Condens. Matter {\bf 9}%
, 693 (1997).

\bibitem{20}  A. J. Leggett, in {\it Quantum Tunneling of Magnetization},
edited by L. Gunther and B. Barbara (Kluwer, Dordrecht, 1995).

\bibitem{21}  A. Garg, J. Appl. Phys. {\bf 76}, 6168 (1994); Phys. Rev.
Lett. {\bf 74}, 1458 (1995).

\bibitem{22}  R. Sessoli, D. Gatteschi, A. Caneschi, M. A. Novak, Nature 
{\bf 365}, 141 (1993); C. Paulsen and J. -G. Park, in {\it Quantum Tunneling
of Magnetization}, edited by L. Gunther and B. Barbara (Kluwer, Dordrecht,
1995); M. A. Novak and R. Sessoli, in {\it Quantum Tunneling of Magnetization%
}, edited by L. Gunther and B. Barbara (Kluwer, Dordrecht, 1995); J. M.
Hern\'{a}ndez, X. X. Zhang, F. Luis, J. Bartolom\'{e}, J. Tejada, and R.
Ziolo, Europhys. Lett. {\bf 35}, 301 (1996); L. Thomas, F. Lionti, R.
Ballou, D. Gatteschi, R. Sessoli, and B. Barbara, Nature (London) {\bf 383},
145 (1996); J. R. Friedman, M. P. Sarachik, J. Tejada, and R. Ziolo, Phys.
Rev. Lett. {\bf 76}, 3820 (1996); J. M. Hern\'{a}ndez, X. X. Zhang, F. Luis,
J. Tejada, J. R. Friedman, M. P. Sarachik, and R. Ziolo, Phys. Rev. B {\bf 55%
}, 5858 (1997); F. Lionti, L. Thomas, R. Ballou, B. Barbara, A. Sulpice, R.
Sessoli, and D. Gatteschi, J. Appl. Phys. {\bf 81}, 4608 (1997); D. A.
Garanin and E. M. Chudnovsky, Phys. Rev. B {\bf 56}, 11102 (1997).

\bibitem{23}  A.-L. Barra, P. Debrunner, D. Gatteschi, C. E. Schulz, R.
Sessoli, Europhys. Lett. {\bf 35}, 133 (1996); C. Sangregorio, T. Ohm, C.
Paulsen, R. Sessoli, and D. Gatteschi, Phys. Rev. Lett. {\bf 78}, 4645
(1997); W. Wernsdorfer and R. Sessoli, Science {\bf 284}\-, 133 (1999).
\end{references}
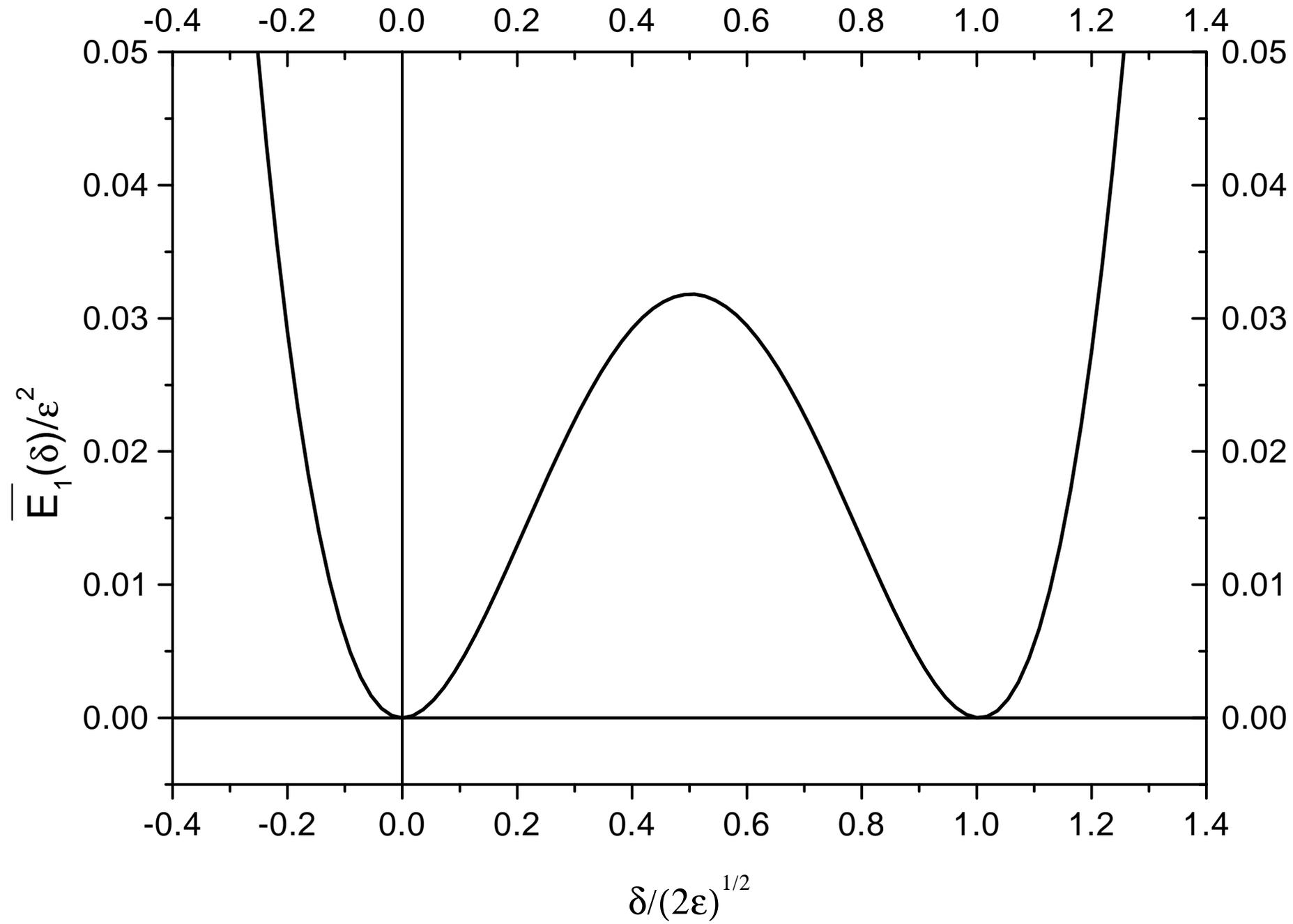
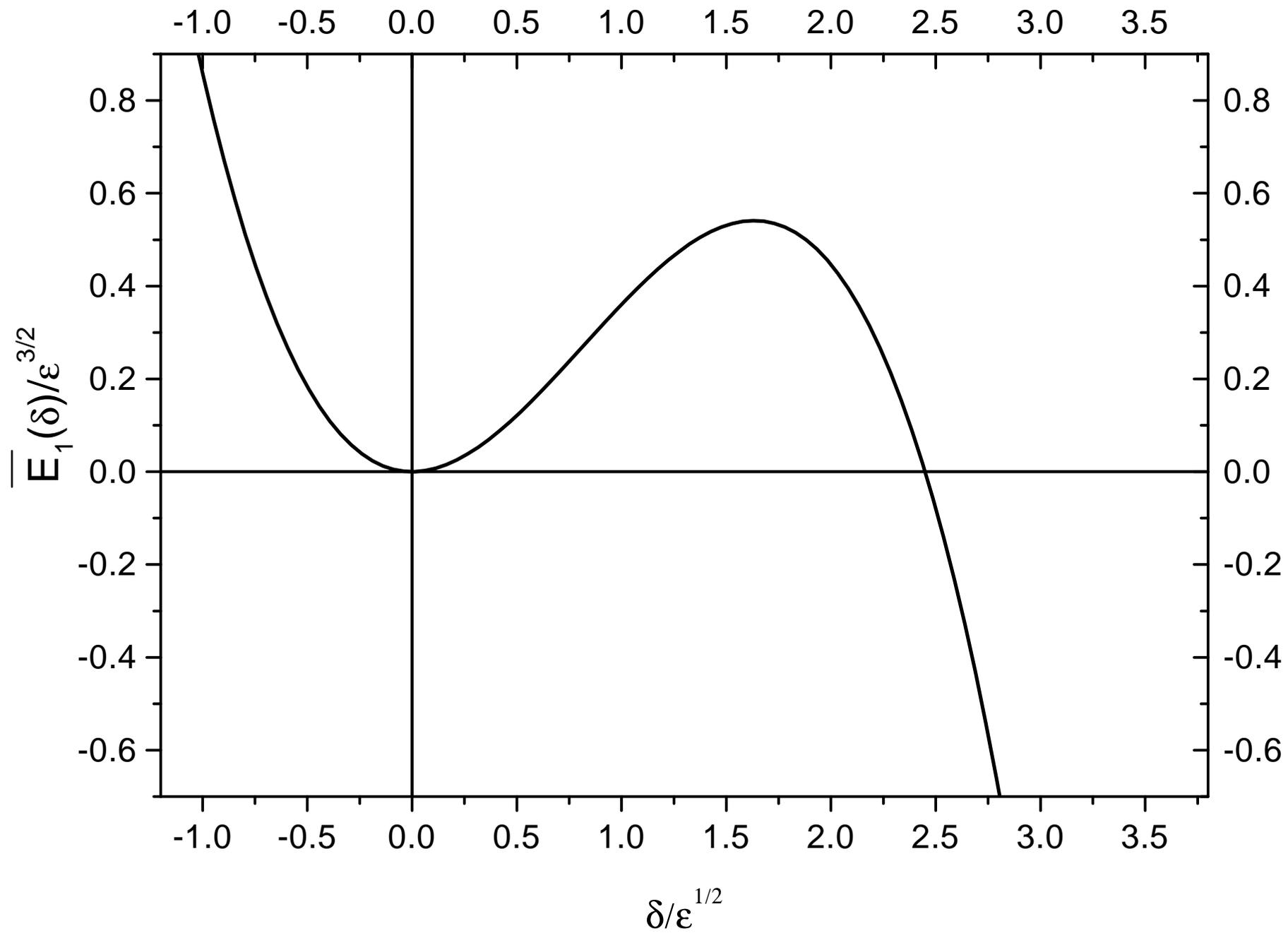
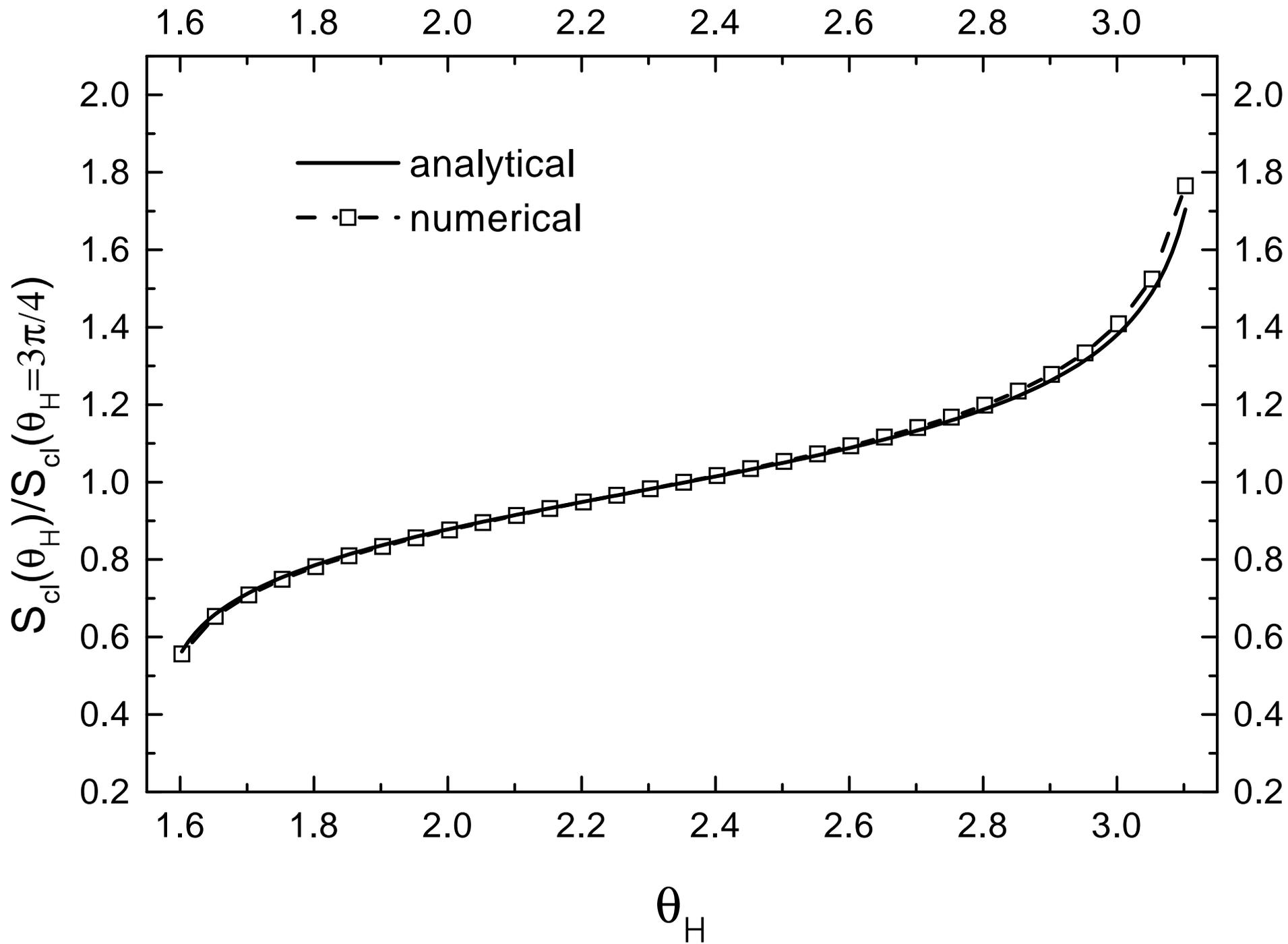
\end{document}